\documentclass[twocolumn]{aastex62}
\usepackage{lipsum}

\graphicspath{{./}{figures/}}

\received{August 28, 2018}
\revised{August 29, 2018}
\accepted{\today}
\submitjournal{ApJ}

\shorttitle{AGN FEEDBACK IN MACS J1447.4+0827}
\shortauthors{Prasow-\'{E}mond et al.}

\begin{document}

\title{A MULTIWAVELENGTH STUDY OF THE MASSIVE COOL CORE CLUSTER MACS J1447.4+0827}

\author[0000-0002-0786-7307]{M. Prasow-\'{E}mond}
\altaffiliation{Email: myriampe@astro.umontreal.ca}
\affil{D\'{e}partement de Physique, Universit\'{e} de Montr\'{e}al, C.P. 6128, Succ. Centre-Ville, Montr\'{e}al, QC H3C 3J7, Canada}

\author{J. Hlavacek-Larrondo}
\affil{D\'{e}partement de Physique, Universit\'{e} de Montr\'{e}al, C.P. 6128, Succ. Centre-Ville, Montr\'{e}al, QC H3C 3J7, Canada}

\author[0000-0003-2001-1076]{C. L. Rhea}
\affil{D\'{e}partement de Physique, Universit\'{e} de Montr\'{e}al, C.P. 6128, Succ. Centre-Ville, Montr\'{e}al, QC H3C 3J7, Canada}

\author{M. Latulippe}
\affil{D\'{e}partement de Physique, Universit\'{e} de Montr\'{e}al, C.P. 6128, Succ. Centre-Ville, Montr\'{e}al, QC H3C 3J7, Canada}

\author{M.-L. Gendron-Marsolais}
\affil{D\'{e}partement de Physique, Universit\'{e} de Montr\'{e}al, C.P. 6128, Succ. Centre-Ville, Montr\'{e}al, QC H3C 3J7, Canada}
\affil{European Southern Observatory, Alonso de C\'{o}rdova 3107, Vitacura, Casilla 19001, Santiago, Chile}

\author{A. Richard-Laferri\`{e}re}
\affil{D\'{e}partement de Physique, Universit\'{e} de Montr\'{e}al, C.P. 6128, Succ. Centre-Ville, Montr\'{e}al, QC H3C 3J7, Canada}
\affil{Institute of Astronomy, University of Cambridge, Madingley Road, Cambridge, CB3 0HA, UK}

\author{J. S. Sanders}
\affil{Max-Planck-Institut f\"{u}r extraterrestrische Physik, 85748 Garching, Germany}

\author{A. C. Edge}
\affil{Institute of Computational Cosmology, Department of Physics, Durham University, Durham, DH1 3LE, UK}

\author{S. W. Allen}
\affil{Department of Physics, Stanford University, 452 LomitaMall, Stanford, CA 94305-4085, USA}
\affil{Kavli Institute for Particle Astrophysics and Cosmology, Stanford University, 382 Via Pueblo Mall, Stanford, CA 94305-4060, USA}

\author{A. Mantz}
\affil{Department of Physics, Stanford University, 452 LomitaMall, Stanford, CA 94305-4085, USA}
\affil{Kavli Institute for Particle Astrophysics and Cosmology, Stanford University, 382 Via Pueblo Mall, Stanford, CA 94305-4060, USA}

\author{A. von der Linden}
\affil{Department of Physics and Astronomy, Stony Brook University, Stony Brook, NY 11794, USA}

\noaffiliation



\begin{abstract}
Clusters of galaxies are outstanding laboratories for understanding the physics of supermassive black hole feedback. Here, we present the first \textit{Chandra}, Karl G. Janksy Very Large Array and \textit{Hubble Space Telescope} analysis of MACS J1447.4+0827 ($z = 0.3755$), one of the strongest cool core clusters known, in which extreme feedback from its central supermassive black hole is needed to prevent the hot intracluster gas from cooling. Using this multiwavelength approach, including 70 ks of \textit{Chandra} X-ray observations, we detect the presence of collimated jetted-outflows that coincides with a southern and a northern X-ray cavity. The total mechanical power associated with these outflows ($P_{\mathrm{cav}} \approx 6 \times 10^{44}$ erg s$^{-1}$) is roughly consistent with the energy required to prevent catastrophic cooling of the hot intracluster gas ($L_{\mathrm{cool}} = 1.71 \pm 0.01 \times 10^{45}$ erg s$^{-1}$ for t$_\mathrm{cool}$ = 7.7 Gyrs); implying that powerful supermassive black hole feedback has been in place several Giga-years ago in MACS J1447.7+0827. In addition, we detect the presence of a radio mini-halo that extends over 300 kpc in diameter ($P_{1.4 \mathrm{GHz}} = 3.0 \pm 0.3 \times 10^{24}$ W Hz$^{-1}$). The X-ray observations also reveal a $\sim20$ kpc plume-like structure that coincides with optical dusty filaments that surround the central galaxy. Overall, this study demonstrates that the various physical phenomena occurring in the most nearby clusters of galaxies are also occurring in their more distant analogues.
\end{abstract}

\keywords{Galaxies: clusters: individual (MACS J1447.4+0827) --- X-rays: galaxies: clusters --- black hole physics --- galaxies: jets}


\section{Introduction} \label{sec:intro}
\begin{figure*}
\includegraphics[width=.515\textwidth]{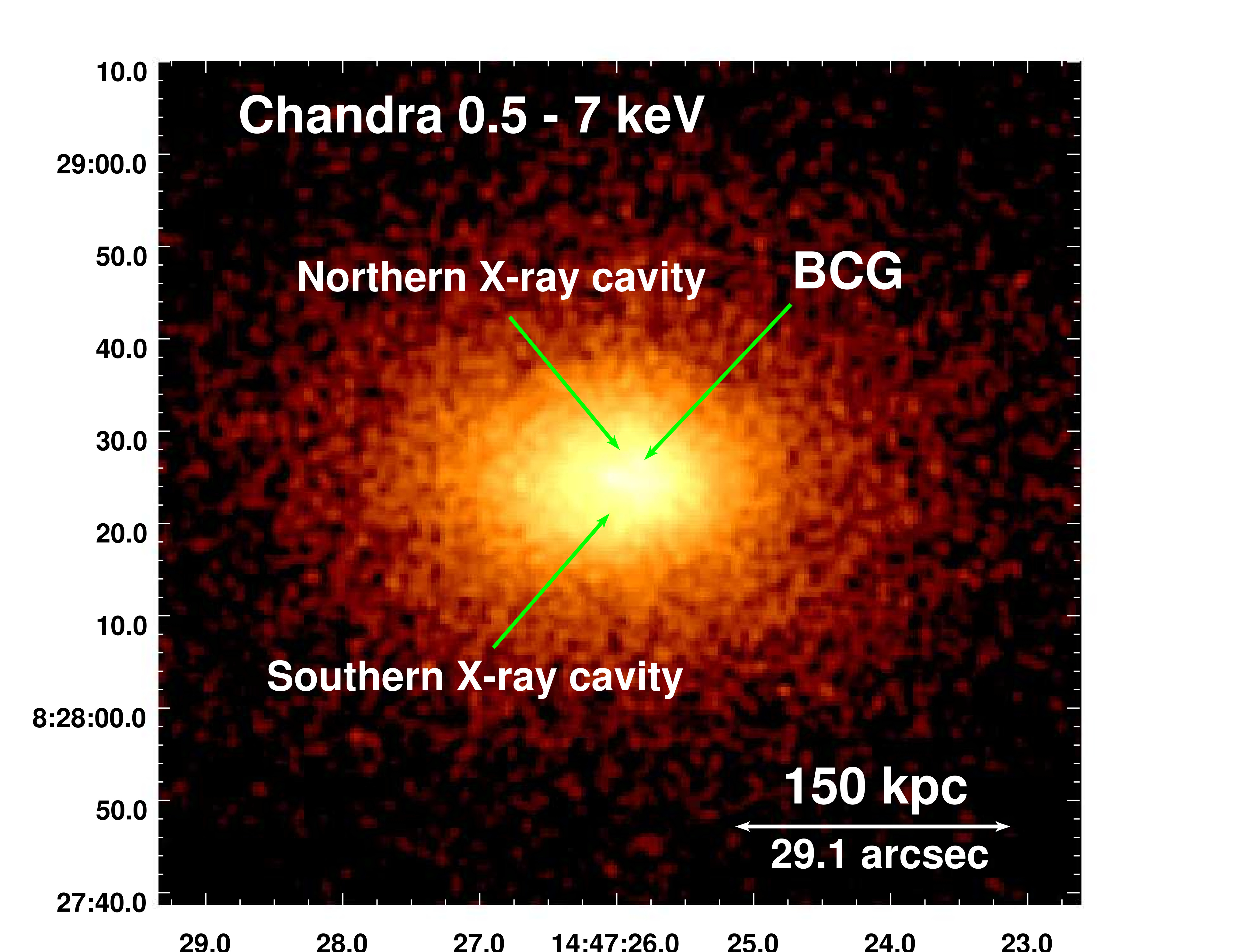}
\includegraphics[width=.515\textwidth]{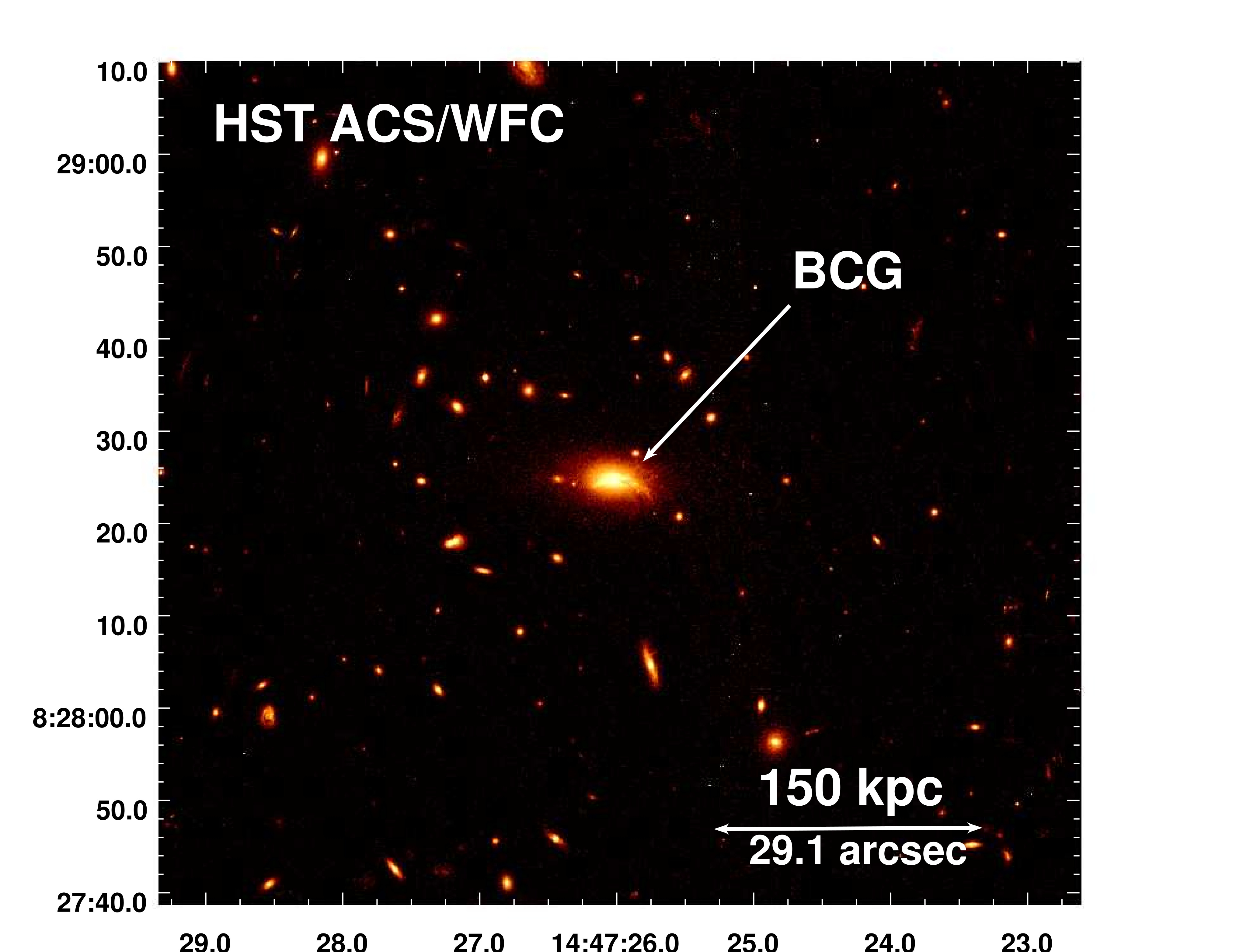}
\caption{\label{fig:CXO_HST}{Left -- Cleaned, exposure-corrected and background subtracted 0.5 -- 7 keV \textit{Chandra} image of MACS J1447.4 + 0827. The total exposure time is 76.3 ks and the image has been smoothed with a $\sigma = $ 2 gaussian function. We highlighted the position of two X-ray cavities, although they are not visible in the raw image, for reference (see Section \ref{sec:tcool} for evidence of their existence). The location of the BCG is also shown. Right -- Same-scale \textsc{HST} ACS/WFC F814W image centered on the BCG.}}
\end{figure*}

Clusters of galaxies are the largest gravitationally bound structures in the Universe. They are primarily composed of dark matter and an intracluster medium (ICM). They also contain hundreds to thousands of galaxies, including some of the most massive galaxies in the universe. At the center, a giant galaxy usually stands out, known as the Brightest Cluster Galaxy (BCG). The BCG is generally an elliptical galaxy hosting, at its center, a supermassive black hole (SMBH) that can generate powerful outflows jets (e.g. \citealt{mcconnell_two_2011}).

We generally classify clusters into two categories based on the spatial distribution of the hot ICM: cool core clusters, where the X-ray surface brightness distribution peaks steeply towards the center and the temperature drops abruptly in the core by a factor of $\sim$ 2-3, and non-cool core clusters in which the X-ray surface brightness distribution is less peaked and more uniform. In the centers of cool core clusters, the optically thin, almost completely ionized ICM has a radiative cooling time that is remarkably short ($\ll$ 10$^{9}$ yr; e.g. \citealt{voigt_thermal_2004, peterson_x-ray_2006, hlavacek-larrondo_hunt_2012}). Left to its own devices, this gas would be forecasted to cool and form stars at rates of several hundred to thousands of solar masses per year (e.g. \citealt{fabian_cooling_1994}). However, the ICM is not cooling as fast as predicted and therefore the observed star formation rates are much lower than expected (e.g. \citealt{johnstone_optical_1987,nulsen_star_1987,odea_infrared_2008,rafferty_regulation_2008,gaspari_solving_2013,mcdonald_revisiting_2018}).

This discrepancy is likely due to the Active Galactic Nucleus (AGN) located at the center of the BCG through a mechanism known as AGN feedback. According to this mechanism, the AGN inflates cavities (or bubbles) in the ICM, detectable using X-ray emission, through relativistic jetted outflows visible at radio wavelengths. The exact mechanism that reheats uniformly the ICM is not yet known, although shock, sound waves and turbulent heating are likely part of the process (e.g. \citealt{birzan_systematic_2004,birzan_radiative_2010, dunn_investigating_2006,dunn_investigating_2008, nulsen_radio_2009, cavagnolo_relationship_2010, dong_prevalence_2010, hlavacek-larrondo_hunt_2012,zhuravleva_turbulent_2014,zhuravleva_nature_2016,zhuravleva_gas_2018}). 

The goal of this paper is to better understand these AGN processes by targetting one of the most massive and strongest cool core clusters known, in which feedback must be operating in an extreme level to offset cooling. 
\begin{figure*}
\begin{center}
    \includegraphics[width=0.32\textwidth]{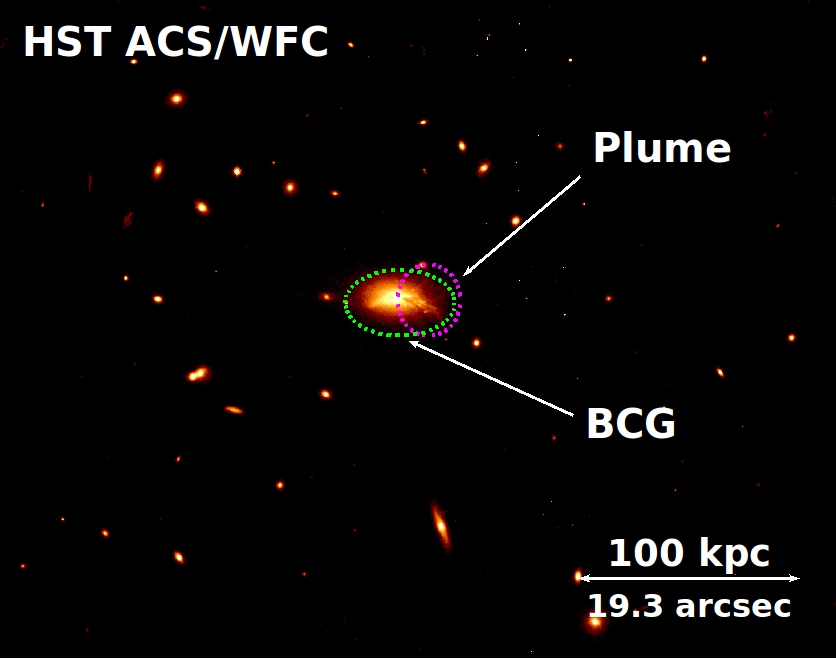}
    \includegraphics[width=0.32\textwidth]{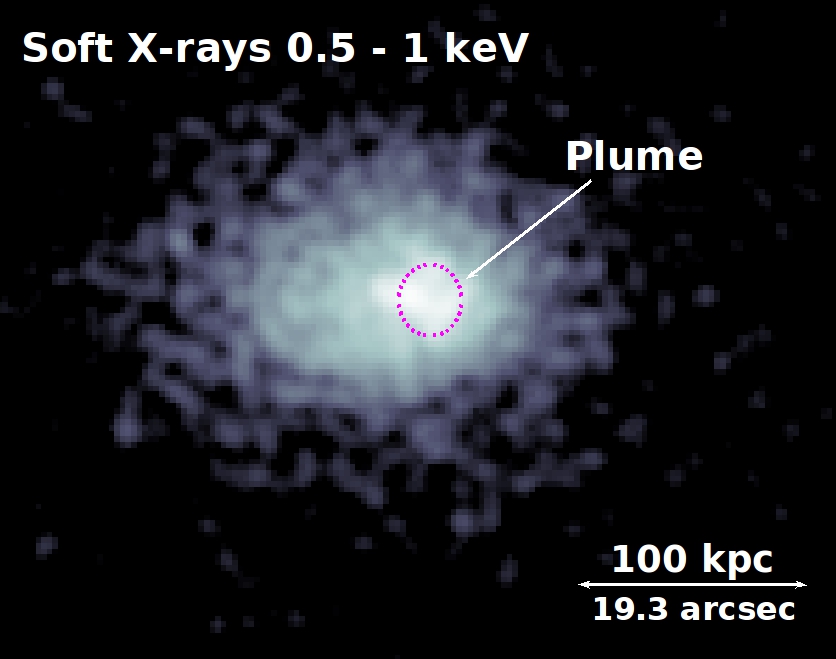}
    \includegraphics[width=0.32\textwidth]{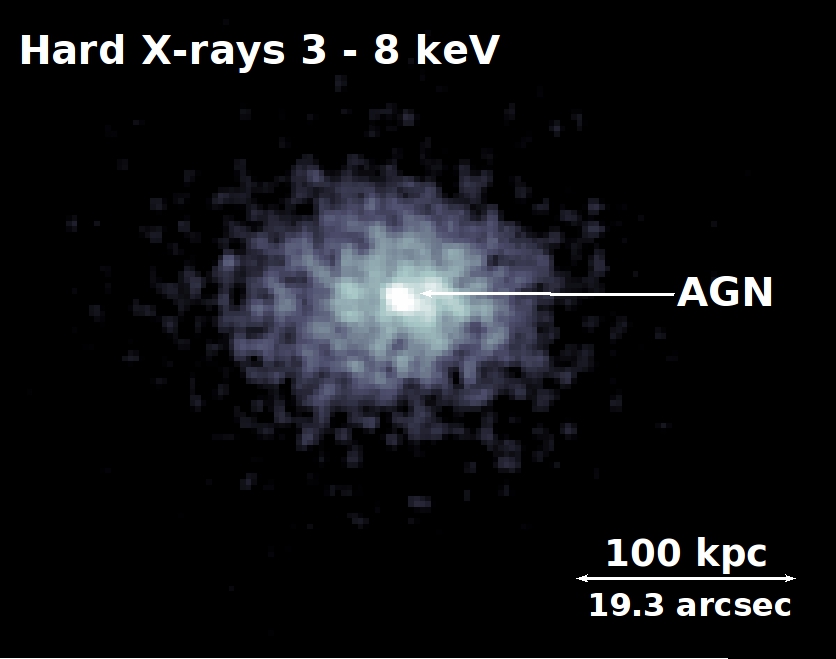}
    \caption{\label{fig:HARDSOFT}{Left -- \textsc{HST} ACS/WFC F814W image centered on the BCG, same as the right panel of Fig. \ref{fig:CXO_HST}. The BCG has been circled by a green ellipse and the plume structure by a magenta ellipse. Middle -- Soft 0.5 -- 1 keV \textit{Chandra} image of MACS J1447.4+0827. The image has been smoothed with a $\sigma = $ 2 gaussian function, and this is the kernel that best highlights the feature. The plume-like structure is indicated by the same magenta ellipse. Right -- Same-scale hard 3 -- 8 keV \textit{Chandra} image, where the AGN location is indicated. The image has been smoothed with a $\sigma = $ 2 gaussian function.}}
\end{center}
\end{figure*}
\textit{The MAssive Cluster Survey} (MACS) consists of a sample of about 120 very X-ray luminous clusters located at a redshift between 0.3 and 0.7 (e.g. \citealt{ebeling_macs:_2001,ebeling_complete_2007,ebeling_x-ray_2010}). The clusters have been selected from the ROSAT All-Sky Survey and were spectroscopically confirmed. The MACS survey has allowed the discovery of new massive clusters at $z >$ 0.3 and thus improved the ability to study their evolution as well as the physical and cosmological parameters driving them. MACS clusters have also been the focus of several gravitational lensing surveys (e.g. MACS J0416.1-2403, \citealt{kaurov_highly_2019}; MACS J0717.5+3745, \citealt{zitrin_largest_2009,jauzac_weak_2012,limousin_strong_2012}; MACS J1149.5+2223, \citealt{smith_hubble_2009}; MACS J0416.1-2403, \citealt{chirivi_macs_2018}; Weighing the Giants program, \citealt{von_der_linden_robust_2014,applegate_weighing_2014,applegate_cosmology_2016}).


In this paper, we present a multiwavelength study of MACS J1447.4+0827 ($z$ = 0.3755), one of the strongest and most X-ray luminous cool core clusters in the MACS survey, similar to well-studied systems such as RBS 797 (e.g. \citealt{schindler_discovery_2001,de_filippis_recent_2002,jetzer_morphology_2002,castillo-morales_clusters_2003,gitti_multifrequency_2006}), MACS J1931.8-2634 (e.g. \citealt{ehlert_extreme_2011,santos_starbursting_2016}), IRAS 09104+4109 (e.g. \citealt{kleinmann_properties_1988,hines_highly_1993,fabian_massive_1999,franceschini_bepposax_2000,iwasawa_chandra_2001,piconcelli_xmm-newton_2007,vignali_nature_2011,farrah_geometry_2016}) and RX J1532.9+3021 (e.g. \citealt{hlavacek-larrondo_probing_2013,gupta_radial_2016}). Here, we present the first detailed study of this cluster. This paper is based on new \textit{Chandra X-ray Observatory} and Karl G. Janksy Very Large Array (\textsc{JVLA}) observations of the source and focuses on understanding its feedback properties. \cite{richard} recently presented the radio mini-halo properties of this cluster.

Section \ref{sec:reduction} introduces the X-ray, radio and optical observations and the data reduction and processing. Section \ref{sec:tcool} presents the X-ray imaging analysis of the cluster. Section \ref{sec:xanalysis} presents a detailed X-ray spectral analysis of the thermodynamic properties of MACS J1447.4+0827. In Section \ref{sec:radiostruc}, we present the radio analysis of the cluster. In Section \ref{sec:agn}, we highlight the properties of the central AGN. Finally, in Section \ref{sec:discussion}, we discuss our results and their implications.

We use $H_0$ = 70 km s$^{-1}$ Mpc$^{-1}$, $\Omega_\mathrm{m}$ = 0.3 and $\Omega_\mathrm{\Lambda}$ = 0.7 as values of the cosmological parameters, and $z$ = 0.3755 as the cluster redshift (1$"$ = 5.157 kpc; \citealt{ebeling_x-ray_2010}). Errors are 1$\sigma$ (68.3\%) throughout the paper (unless mentioned).

\section{Observations and data reduction} \label{sec:reduction}
\subsection{X-rays: Chandra observations} \label{sec:chandra}

The \textit{Chandra} Advanced CCD Imaging Spectrometer (ACIS) in VFAINT mode was used to observe MACS J1447.4+0827 three times: first in 2008, centered on the ACIS-S3 back-illuminated chip (12 ks; ObsID 10481; PI Hicks) and then twice in 2016, centered on the ACIS-I3 front-illuminated chip (43 ks; ObsID 17233 and 25 ks; ObsID 18825; PI Hlavacek-Larrondo). These observations are presented in Table \ref{tab1}.

\begin{table}[ht]
\begin{center}
\caption{\label{tab1}{\textit{Chandra} observations.}}
    \begin{tabular}{cccc}
    \hline\hline\\[-1.em]
    Observation number & Date & Detector & Exposure [ks] \\
    \hline
    10481 & 12/14/2008 & ACIS-S & 11.1  \\
    17233 & 04/05/2016 & ACIS-I & 41.0  \\
    18825 & 04/06/2016 & ACIS-I & 24.2  \\
    \hline
    \end{tabular}
\end{center}
\end{table}

Starting from the event 1 file, the ObsIDs were cleaned, processed and calibrated using the \texttt{CIAO} software (\texttt{CIAO v4.10} and \texttt{CALDB v4.7.8}). After reprocessing the data with the \texttt{chandra$\_$repro} routine, we excluded 9 point sources by eye using \texttt{ds9} and we removed flares with \texttt{deflare} with \texttt{lc$\_$clean} and a 3$\sigma$ threshold. The net exposure times are presented in Table \ref{tab1}. 

\begin{table*}
\centering
\caption{\label{tab2}{\textsc{JVLA} Radio observations.}}
	\begin{tabular}{cccccc}
    \hline\hline\\[-1.em]
	Date & Frequency [GHz] & Bandwidth [GHz] & Configuration & Duration [min] & $\sigma_{\mathrm{rms}}$ [mJy/beam] \\
	\hline
	08/14/2015 & 1.5 & 1 & A & 180 & 0.011\\
	05/25/2016 & 1.5 & 1 & B & 180 & 0.016\\
	02/02/2016 & 1.5 & 1 & C & 180 & 0.015\\
	\hline
	\end{tabular}
\end{table*}
Background event files, background images and background-subtracted images were generated using \texttt{blanksky} and \texttt{blanksky$\_$image}. Finally, the tool \texttt{merge$\_$obs} was applied to the observations to merge the event files, which created a final exposure-corrected image. The cleaned, merged, background subtracted and exposure-corrected \textit{Chandra} image is presented in the left panel of Fig. \ref{fig:CXO_HST}. We note that we also computed several of the cluster properties using a background region far from the cluster on the ACIS-I chips and obtained consistent results with the \texttt{blanksky} background event files.

\texttt{XSPEC} (\texttt{v12.10.0c}) was used to analyze spectroscopically the X-ray observations of MACS J1447.4+0827. Throughout this study, we used the \cite{anders_abundances_1989} abundance ratios and the \cite{kalberla_leiden/argentine/bonn_2005} value of the Galactic absorption ($n_\mathrm{h}=0.0224 \times10^{22}$ cm$^{-2}$). We verified this value by picking a certain region located between 50 kpc and 200 kpc from the center -- in which there is no emission (e.g. from the AGN) -- and fitting an absorbed \texttt{apec} plasma model to the spectrum in the 0.5 -- 7 keV energy range while letting the Galactic absorption, temperature, abundance and normalization parameters free to vary. We obtained results consistent with the \cite{kalberla_leiden/argentine/bonn_2005} value within 1$\sigma$.

The middle and right panels of Fig. \ref{fig:HARDSOFT} present the soft 0.5 -- 1 keV and hard 3 -- 8 keV X-ray \textit{Chandra} images. The middle panel reveals a plume-like structure that matches the western filaments of the BCG which we discuss in Section \ref{sec:plume}. The right panel has a bright point-like region in the center coincident with the BCG. We interpret this feature as the central AGN and we present its properties in Section \ref{sec:agn}.

\begin{figure*}
\includegraphics[width=.32\textwidth]{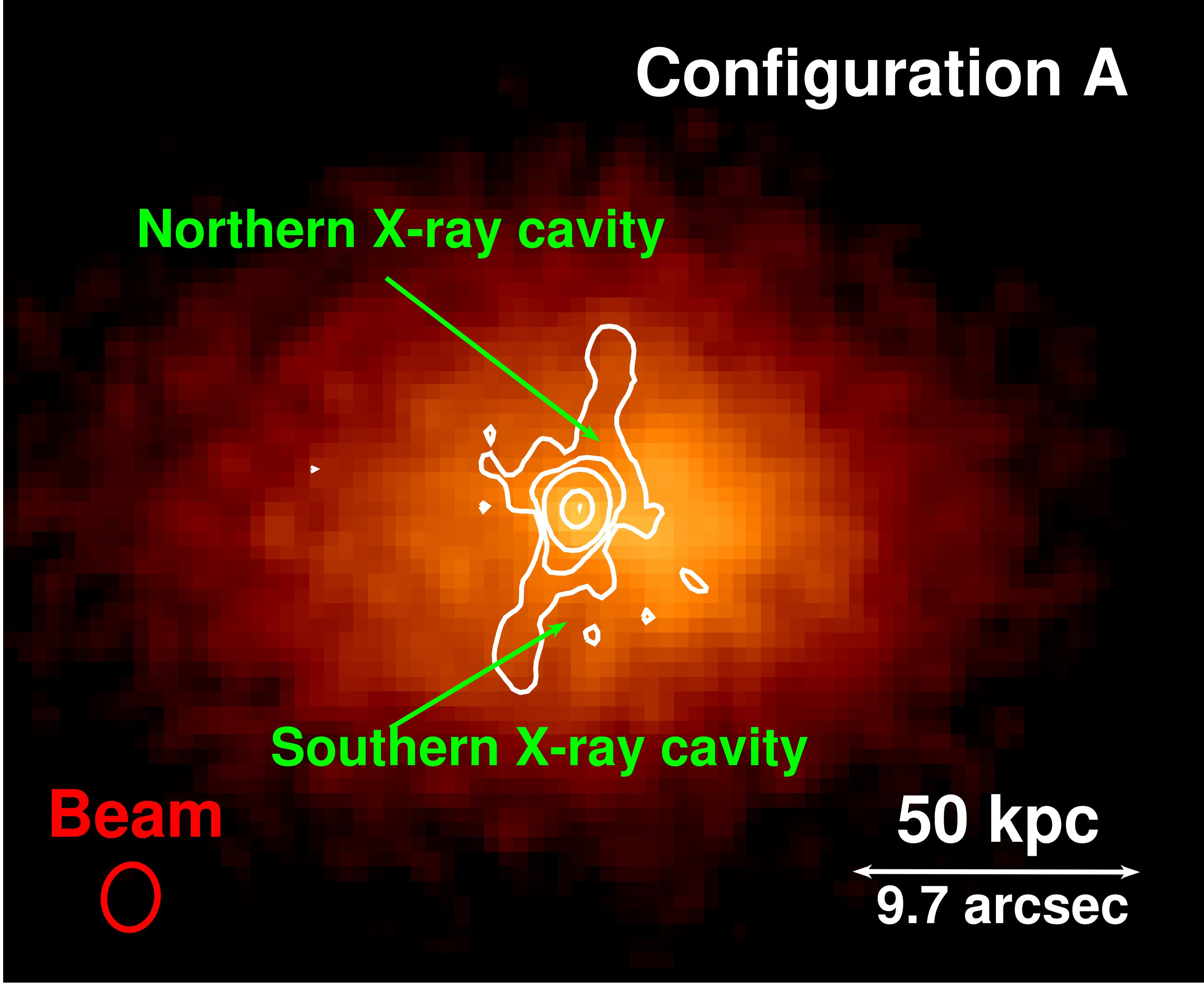}
\includegraphics[width=.32\textwidth]{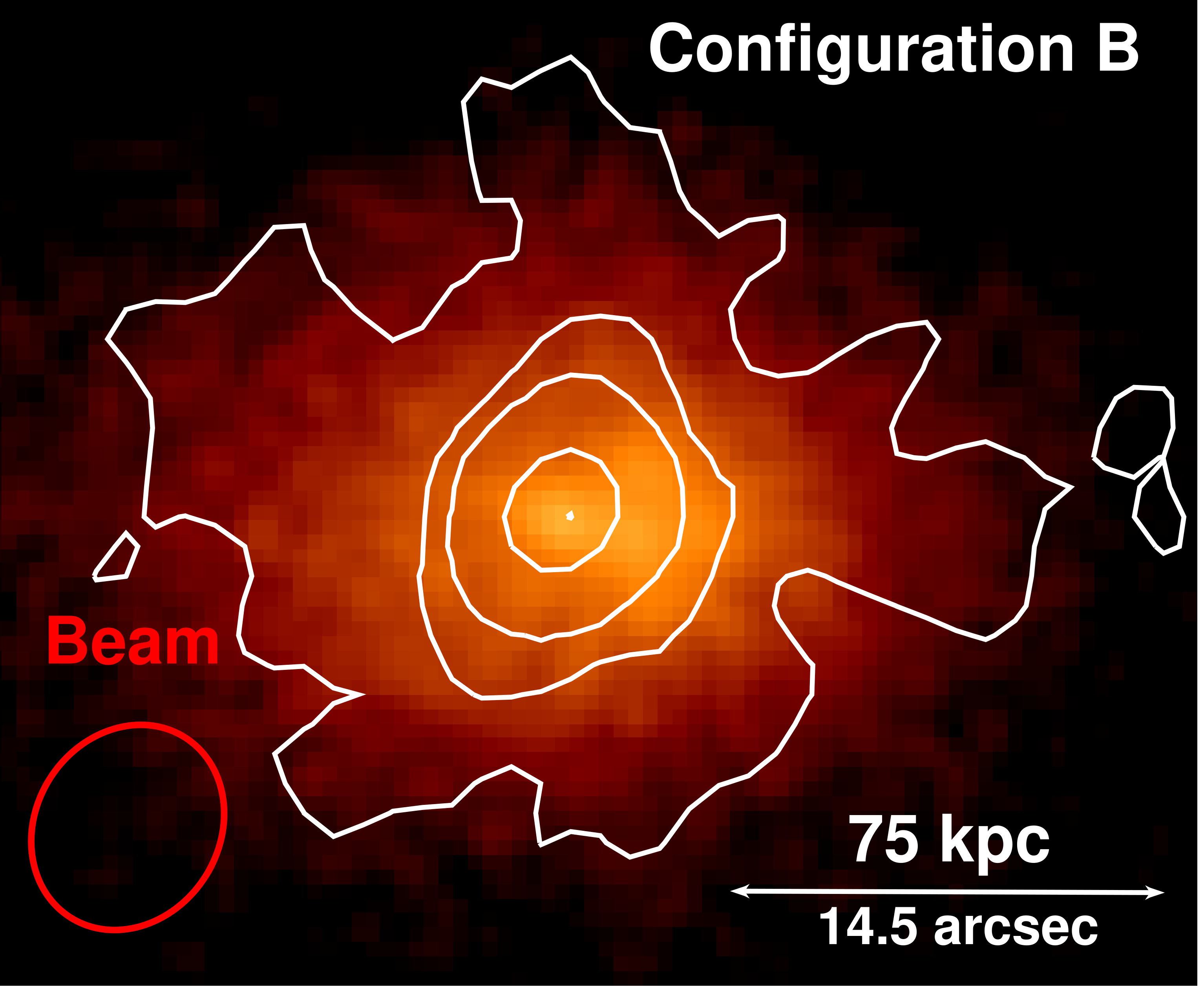}
\includegraphics[width=.32\textwidth]{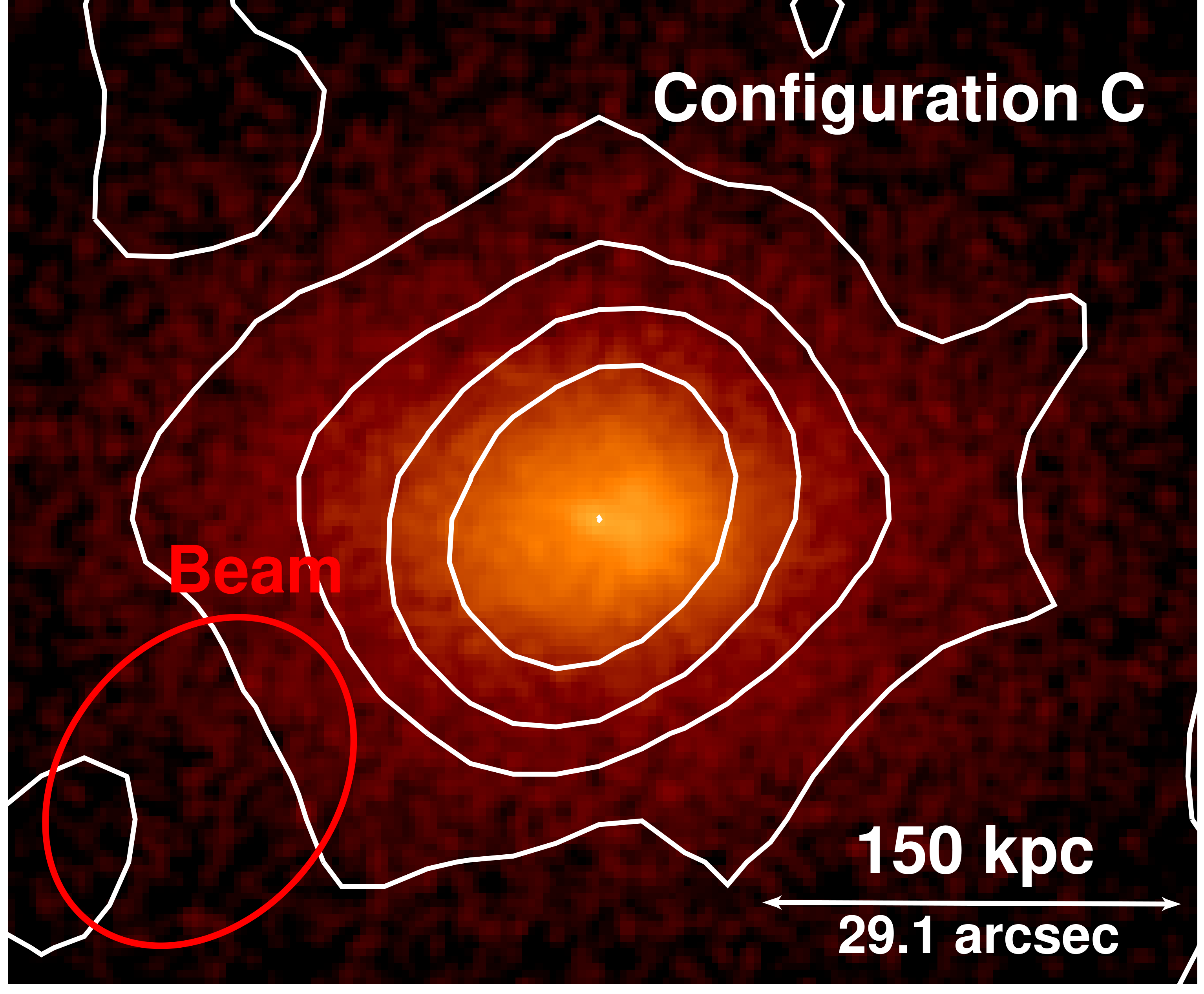}
\caption{\label{fig:radio_cont}{Left -- Background-subtracted and exposure-corrected 0.5 -- 7 keV \textit{Chandra} image (same as Fig. \ref{fig:CXO_HST}) with 1.4 GHz \textsc{JVLA} contours of the configuration A observations. The contours are 4$\sigma_{\mathrm{rms}}$ = 0.044, 0.176, 0.704, 5 and 35 mJy/beam. The beam is 0.9$\arcsec \times 1.1\arcsec$ and has a position angle from north to east of $-7.73^{\circ}$. The location of the northern and southern X-ray cavities are indicated for reference (see Section \ref{sec:tcool} for evidence of their existence). Middle -- Same as left, but with contours of the configuration B observations. The contours are 4$\sigma_{\mathrm{rms}}$ = 0.064, 0.3, 1, 5 and 35 mJy/beam. The beam is 3.1$\arcsec \times 3.6\arcsec$ and has a position angle from north to east of $-32.7^{\circ}$. Right -- Same as left, but with contours of the configuration C observations. The contours go from 4$\sigma_{\mathrm{rms}}$ = 0.060, 0.3, 1, 5 and 39 mJy/beam. The beam is 9.6$\arcsec \times 12.5\arcsec$ and has a position angle from north to east of $-37.8^{\circ}$.}}
\end{figure*}

\subsection{Radio: JVLA observations}\label{sec:radio}

MACS J1447.4+0827 was also observed with the \textsc{JVLA} at 1.4 GHz (L band; PI Hlavacek-Larrondo). We present a summary of the data in Table \ref{tab2}. Configurations A, B and C were obtained to maximize the different scales detected: the A-configuration observations are of comparable resolution to \textit{Chandra}, whereas B- and C-configurations allow for better detection of faint extended radio emission. The data were reduced with the software \texttt{CASA} (\texttt{v4.6}). A more detailed description of the data reduction and imaging processes are presented in \cite{richard}.

The exposure-corrected and background-subtracted 0.5 -- 7 keV \textit{Chandra} image along with the 1.4 GHz \textsc{JVLA} configurations A (left), B (middle) and C (right) contours are shown in Fig. \ref{fig:radio_cont}. The images respectively have beam sizes of 0.9$\arcsec$ $\times$ 1.1$\arcsec$, 3.1$\arcsec$ $\times$ 3.6$\arcsec$ and 9.6$\arcsec$ $\times$ 12.5$\arcsec$. The root-mean-square (rms) reached are shown in Table \ref{tab2}.
   
The A-configuration contours reveal the presence of a compact central AGN that coincides with the X-ray point source seen in the right panel of Fig. \ref{fig:HARDSOFT} and located at RA = $14^{\mathrm{h}}47^{\mathrm{m}} 26.022^{\mathrm{s}}$ and DEC = $+8^{\circ}28\arcmin 25.26\arcsec$. It also reveals the presence of north-south oriented jetted outflows that will be analyzed in Section \ref{sec:power}. Configurations B and C reveal faint emission that extends out to $\sim$ 160 kpc of radius. This type of structure has already been seen in other cool core clusters and we interpreted it as a radio mini-halo (e.g. \citealt{mazzotta_radio_2008}, \citealt{richard}). We characterize the properties of the mini-halo in Section \ref{sec:MH}
  
\subsection{Optical: HST observations}\label{sec:hubble}
The \textit{Hubble Space Telescope} (\textsc{HST}) was used to provide optical images of MACS J1447.4+0827 (PI Ebeling). We used two broad-band filters both using the camera Wide Field Channel (WFC)/HRC: F606W ($\lambda$ = 5907 \AA; $\Delta \lambda$ = 2342 \AA) and F814W ($\lambda$ = 8333 \AA; $\Delta \lambda$ = 2511 \AA). These filters contain the redshifted H$\alpha$ emission line (H$\alpha \lambda$6562) and highlight the filamentary structure of the BCG. The right panel of Fig. \ref{fig:CXO_HST} and the left panel of \ref{fig:HARDSOFT} shows the F814W image taken with the Advanced Camera for Surveys (ACS) WFC. 

\begin{figure*}
\begin{center}
\includegraphics[width=0.9\textwidth]{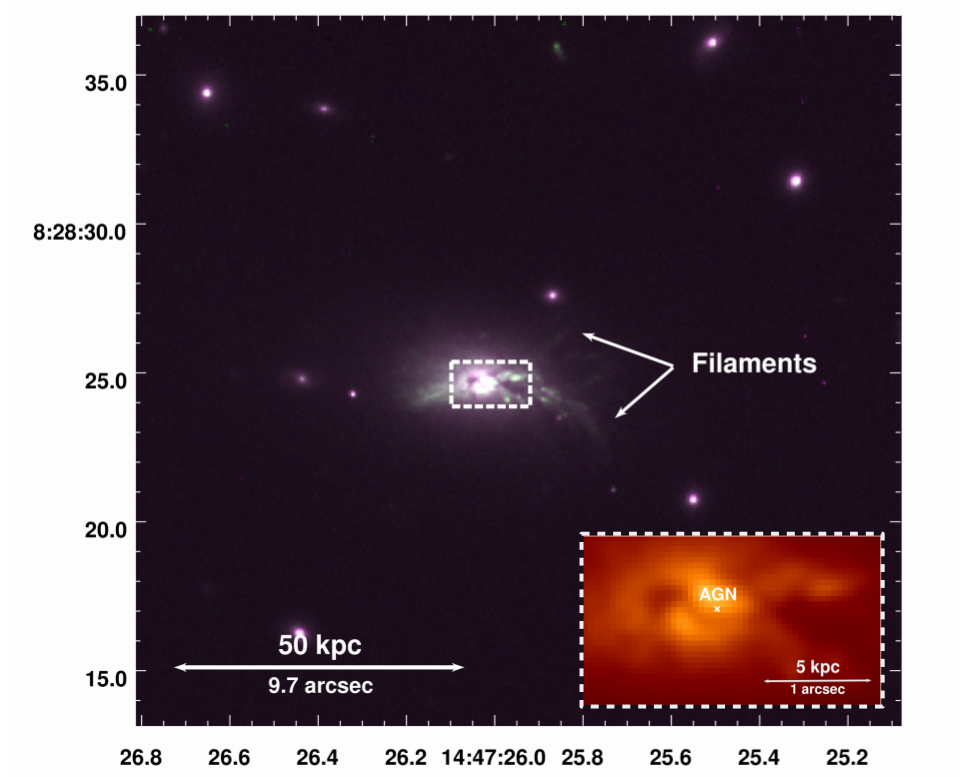}
\caption{\label{fighubble}{Combined \textsc{HST} F814W (red and blue) and F606W (green) images centered on the BCG of MACS J1447.4+0827. The inset to the bottom-right is a zoom on the central part of the galaxy, focusing on the optical nucleus and shows the position of the AGN as detected at 1.4 GHz based on the A-configuration \textsc{JVLA} observations.}}
\end{center}
\end{figure*}

Fig. \ref{fighubble} shows the combined \textsc{HST} F814W (red and blue) and F606W (green) images of MACS J1447.4+0827. The BCG is surrounded by optically bright filaments up to $\sim$ 25 kpc in size. We can denote two prominent filaments: one oriented in the west-south direction and the other in the west-north direction. We analyze the properties of the filaments more in detail in Section \ref{sec:cavities}.


\section{X-ray Imaging analysis of the cluster}\label{sec:tcool}
In the following sections, we present the methods used in this study to highlight the X-ray
cavities. Although the cavities are not clearly visible in the raw image (see Fig. \ref{fig:CXO_HST}), they must exist due to the presence of relativistic outflows (see left panel of Fig. \ref{fig:radio_cont}) that push gas. We therefore applied a total of three different methods to bring out these features.

\subsection{Gaussian Gradient Magnitude}
First, we applied a Gaussian Gradient Magnitude (GGM) filter technique that has recently been applied to astronomical X-ray datasets \citep{sanders_very_2016}. Starting with the exposure-corrected, background-subtracted image of MACS J1447.4+0827, we convolve the image with the multi-directional gradient of a gaussian using the \texttt{scipy} implementation\footnote{The implementation of our code can be found at \url{https://github.com/crhea93/AstronomyTools/tree/master/GGF}}. This process allows us to strategically resolve astrophysical phenomena of different scale lengths by adjusting the gaussian’s sigma value. Here, we have chosen a $\sigma$ = 2 in order to pick out structures corresponding to a 5 kpc scale. We further applied a logarithmic scaling to highlight the complicated substructure. We present the GGM \textit{Chandra} image in the left panel of Fig. \ref{fig:subtracted}. Unfortunately, we do not detect any cavities with this technique, but this image allows to see the variations in the cluster.

\subsection{Unsharp-masked}\label{sec:unsharp}
We then subtracted a 2D gaussian smoothed image ($\sigma$ = 6 pixels) from a less smoothed image ($\sigma$ = 1 pixel) using \texttt{fgauss} and \texttt{farith} from \texttt{CIAO}. This unsharp-masked image is presented in the middle panel of Fig. \ref{fig:subtracted}, where we can denote four cavity-like structures. In the middle of this image, we can also see a brightest part -- a plume-like structure -- that will be described in Section \ref{sec:plume}. We tested several smoothing and binning factors to make the X-ray cavities as clear as possible.

\subsection{Double-$\beta$ model}\label{sec:doublebeta}
To further test the presence of cavities, we used a double-$\beta$ model (e.g. \citealt{jones_structure_1984,vikhlinin_chandra_2005}) to fit the X-ray surface brightness profile, based on the following equation from \cite{cavaliere_x-rays_1976}:
\begin{equation}
\resizebox{0.9\hsize}{!}{$S(r)=S_{01}\Bigg(1+\bigg(\frac{r}{r_{c1}}\bigg)^{2}\Bigg)^{-3\beta_{1}+0.5}+S_{02}\Bigg(1+\bigg(\frac{r}{r_{c2}}\bigg)^{2}\Bigg)^{-3\beta_{2}+0.5}+C$}.
\end{equation}

Here, $S_0$ is the central surface luminosity, $r_c$ is the radius of the core, $\beta$ is the slope and $C$ is the constant associated with the background. 

\begin{figure*}
\begin{center}
\includegraphics[width=.32\textwidth]{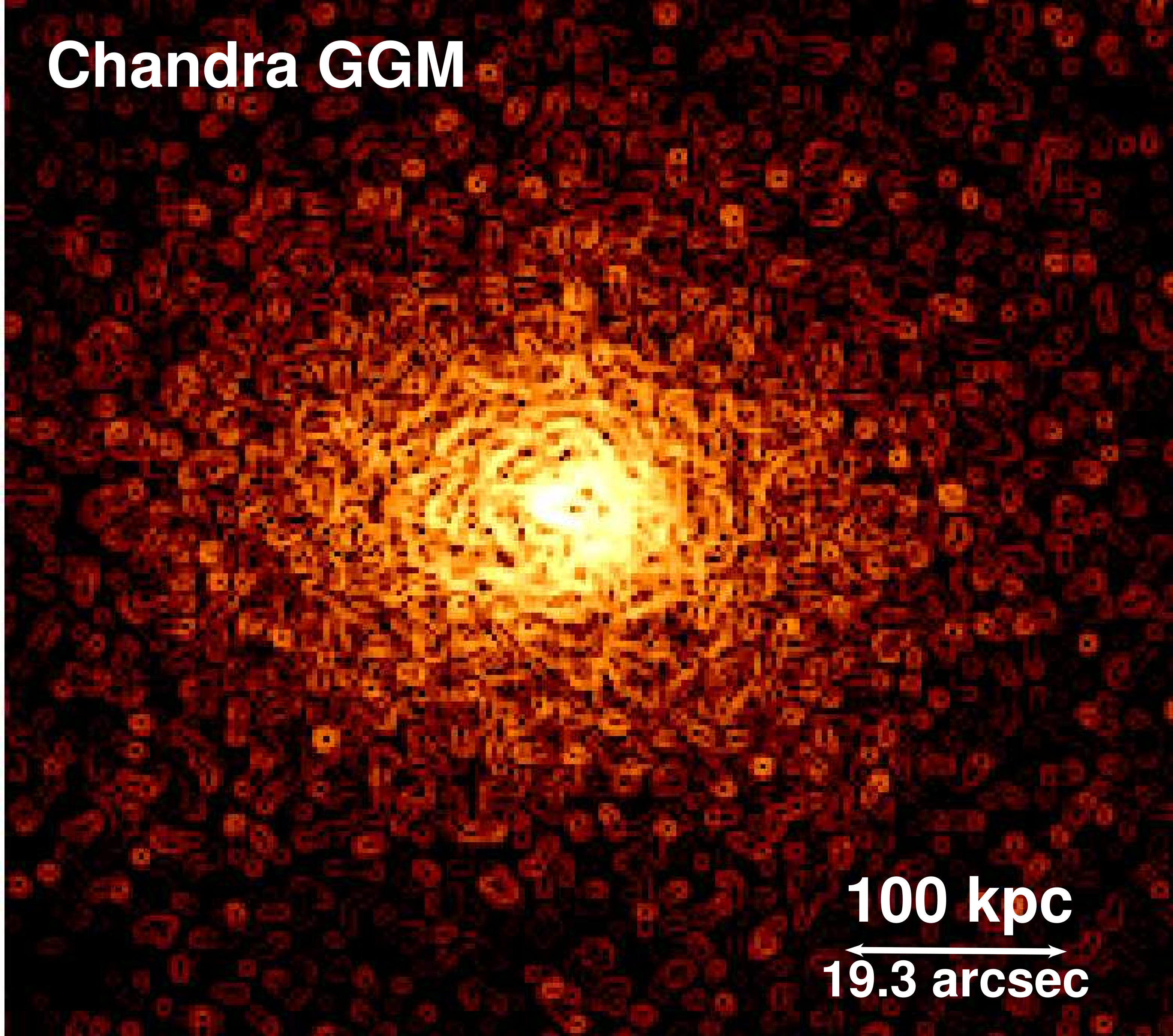}
\includegraphics[width=.32\textwidth]{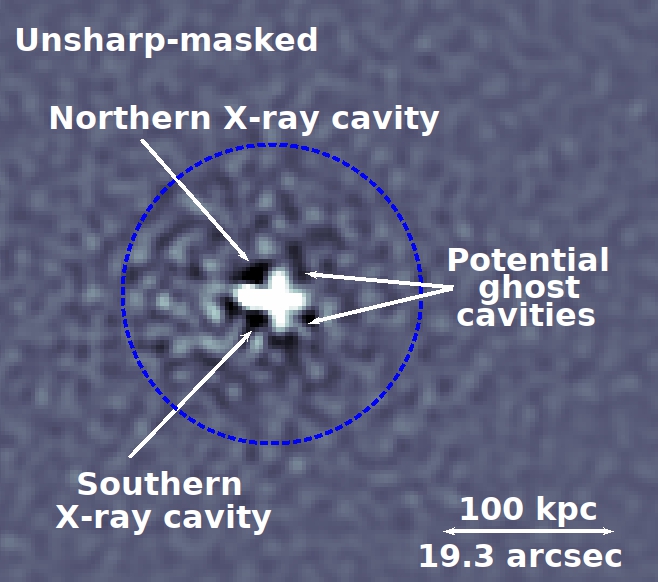}
\includegraphics[width=.29\textwidth]{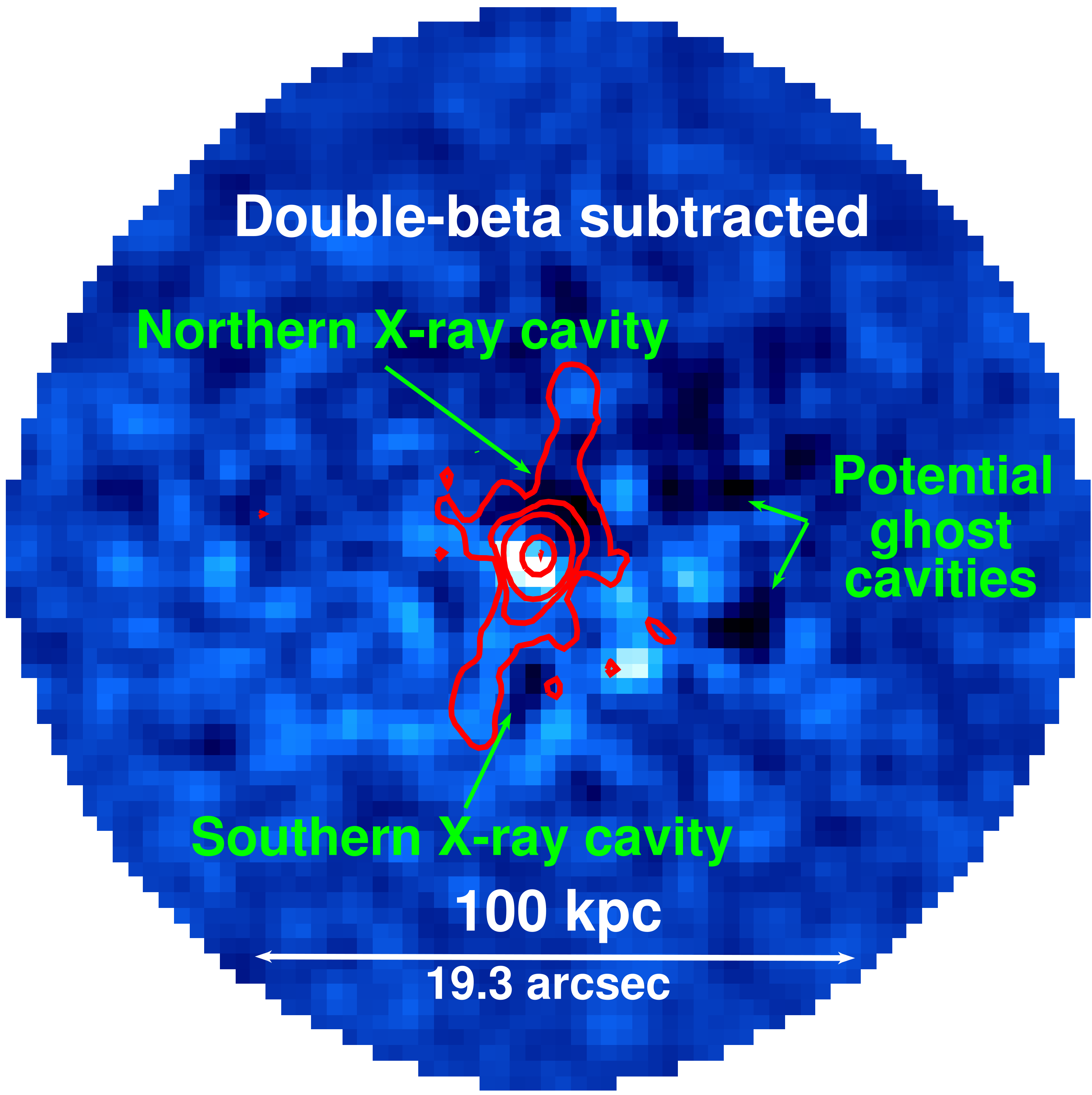}
\caption{\label{fig:subtracted}{Left -- Background-subtracted and exposure-corrected 0.5 -- 7 keV \textit{Chandra} image, same as Fig. \ref{fig:CXO_HST}, but with Gaussian Gradient Magnitude (GGM) filter with $\sigma$ = 2. This image allows us to better detect the complicated substructure. The spirals are caused by point sources. Middle -- Unsharp-masked \textit{Chandra} image where a 2D gaussian image ($\sigma$ = 6) was subtracted from a less smoothed image ($\sigma$ = 1; see Section \ref{sec:unsharp}). Different features are highlighted. The blue circle indicates the size of the double-$\beta$ image (right panel). Right -- Smoothed image (2D gaussian with $\sigma$ = 2), subtracted from a double-$\beta$ model (see Section \ref{sec:doublebeta}). X-ray cavities and potential X-ray ghost cavities are indicated. We added the 1.4 GHz \textsc{JVLA} contours in A-configuration, same as Fig. \ref{fig:radio_cont}.}}
\end{center}
\end{figure*}

\begin{figure*}
\begin{center}
\includegraphics[width=.4\textwidth]{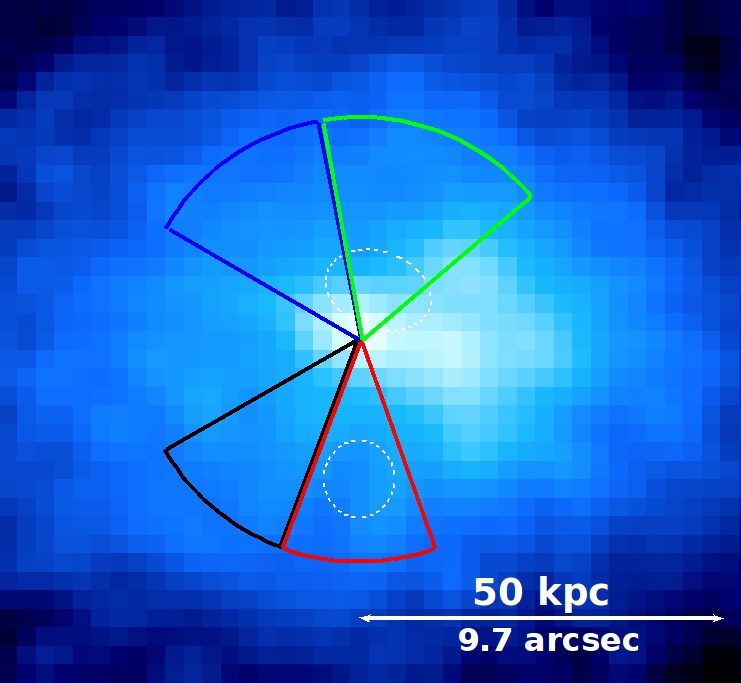}
\includegraphics[width=.55\textwidth]{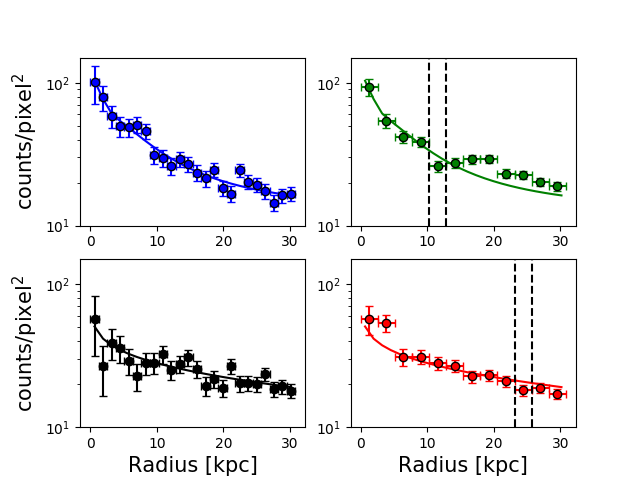}
\caption{\label{fig4dir}{Left -- 0.5 -- 7 keV \textit{Chandra} image of MACS J1447.4+0827 with four sectors overplotted. The white dashed ellipses denote the approximate location of the X-ray cavities. Right -- Double-$\beta$ fitted surface brightness profile along each sector. The dashed lines denote the approximate position of the cavities.}}
\end{center}
\end{figure*}

To create the model, we initially divided a circular region of 100 kpc from the center in 15 equal sectors, and each section was separated in 40 radial ranges. We then extracted the surface brightness of the radial ranges from each sector with the tool \texttt{dmextract} from \texttt{CIAO}. The double-$\beta$ model was fitted to those regions, which produced a brightness profile for every sector. The model was created by assembling the 15 sectors by smoothing over the sharp discontinuities at the edges of the sectors using \texttt{fgauss}. Finally, we subtracted the model image from the original image. We present the results in the right panel of Fig. \ref{fig:subtracted}, in which we added the contours of the \textsc{JVLA} A-configuration observations. Note that we varied the parameters (number of regions, coordinates of the center and the number of regions per sector) and obtained results consistent with the unsharp-masked image in the previous sub-section. Fig. \ref{fig:subtracted} reveals northern and southern potential cavities, but also two potential ghost cavities, consisting of cavities that have been dragging out of the AGN's area to larger radii. To test the presence of the northern and southern cavities, we fitted the double-$\beta$ model along four regions and obtained surface brightness profiles in Fig. \ref{fig4dir}.The northern cavity is detected at 2$\sigma$ and the southern, at 1.7$\sigma$. These detections are statistically weak compare to others clusters (an X-ray cavity surface brightness is normally $\sim$ 20\% below the ICM's), but we can still confirm the presence of two X-ray cavities due to the presence of the relativistic outflows. We further discuss this interesting feature in Section \ref{sec:cavities}.

We note that the 1.4 GHz \textsc{JVLA} emission matches the position of the northern potential cavity, but does not match exactly the southern cavity. Located further away from the AGN, ghost cavities are filled with radio emission with steeper spectrum as their relativistic electrons content got older and are therefore less bright at 1.4 GHz than the inner cavities.
They are also more difficult to see in the X-rays due to the decreasing brightness as a function of the radius. The X-ray energetics of the cavities will be presented in Section \ref{sec:power} and we further discuss the presence of the potential ghost cavities in Section \ref{sec:ghostcav}. 


\section{X-ray spectral analysis of the cluster}\label{sec:xanalysis}

The following sections present the thermodynamic profiles and maps of the cluster.

\subsection{Thermodynamic profiles}\label{sec:thermo}
\begin{figure*}
\includegraphics[width=0.5\textwidth]{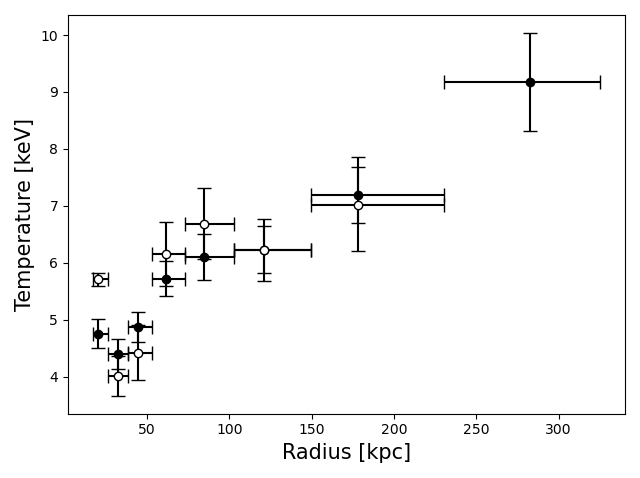}
\includegraphics[width=0.5\textwidth]{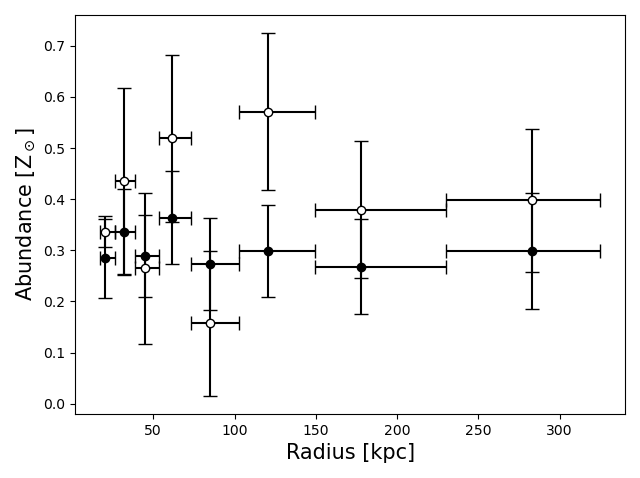}
\includegraphics[width=0.5\textwidth]{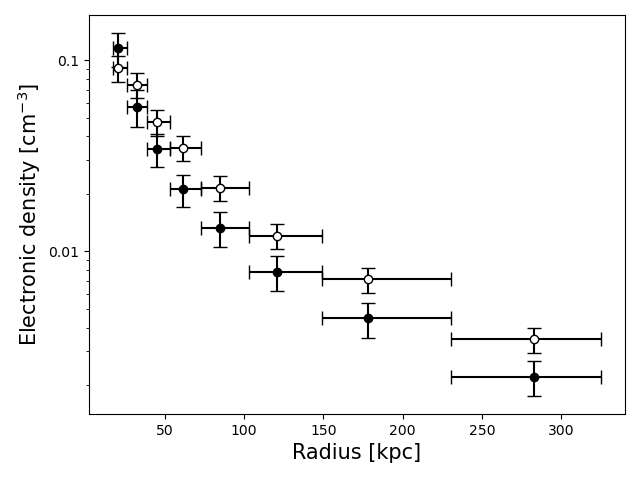}
\includegraphics[width=0.5\textwidth]{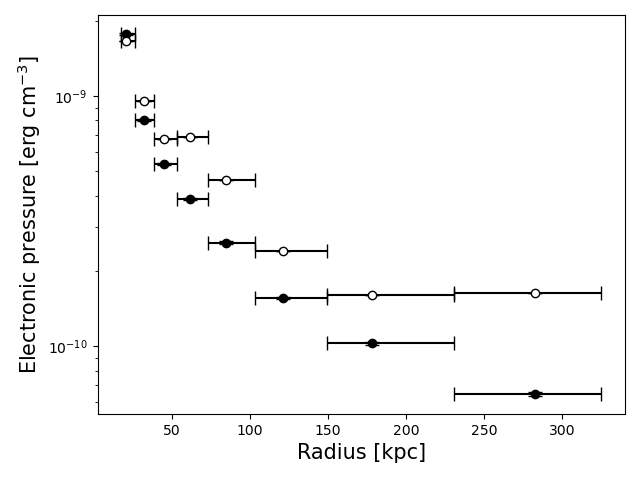}
\caption{\label{figthermo}{Projected (filled symbols) and deprojected (empty symbols) temperature (up-left), abundance (up-right), electron density (down-left) and electron pressure (down-right) profiles using the method described in Section \ref{sec:thermo}. Horizontal error bars indicate the radial extent of the bins used for fitting. Since we here have a elliptical geometry, the radius is equal to the semi-major axis.}}
\end{figure*}

To compute the thermodynamic profiles, we selected nine elliptical regions of $\sim$ 6000 counts with a signal-to-noise ratio of $\sim$ 75 centered on the central AGN from the background-subtracted image. To determine the ellipticity ($\epsilon$) of these regions, we manually traced elliptical regions on the X-ray image on 10 contours smoothed with a $\sigma =$ 10 gaussian function. These regions encompass all the X-ray emission up to a distance of $\sim$ 750 kpc from the center and the ellipticity was defined as the average ($\epsilon$ = 0.26) with a position angle consistent with 0$^{\circ}$. Afterwards, thermodynamics quantities were extracted with \texttt{XSPEC} and fitted with an \texttt{apec} and \texttt{phabs} models. The spectra were then deprojected with the tool \texttt{DSDEPROJ} \citep{sanders_deeper_2007} -- a tool that deprojects the spectra onto 3D ellipsoidal shells \footnote{Details of the code can be found here: \url{https://www-xray.ast.cam.ac.uk/papers/dsdeproj/}}. Note that we varied the number of counts in each region between 3000 and 8000 and obtained similar fits. The projected and deprojected temperature, electron density, electron pressure and abundance profiles are presented in Fig.
\ref{figthermo}. We note that the temperature is higher near the center, which is in agreement with the presence of an AGN. The sudden temperature increase at $\sim$ 50 kpc and $\sim$ 300 kpc can be explained by cold fronts.



Using the deprojected temperature and density profiles, we calculated the core entropy $K_0$, based on the following equation:
\begin{equation}
    K = kT \, n_{\mathrm{e}}^{-2/3}.
\end{equation}

Here, $kT$ is the gas temperature and $n_{\mathrm{e}}$ is the electron density. To calculate $K_0$, we used the deprojected central density of Fig. \ref{figthermo}. The temperature is however strongly influenced by the AGN, so we extrapolated it using the relation $T \propto r^{\mathrm{b}}$, where $r$ is the radius and b $\sim$ 0.3 \citep{voigt_thermal_2004}. We obtained a core entropy of $K_{0}$ = (12 $\pm$ 2) keV cm$^{2}$. According to \cite{cavagnolo_entropy_2008}, this result is another confirmation that MACS J1447.4+0827 is a very strong cool core cluster, with $K_0$ well under 30 keV cm$^{2}$, the threshold for cool core clusters.

\begin{figure}[ht]
\includegraphics[width=\linewidth]{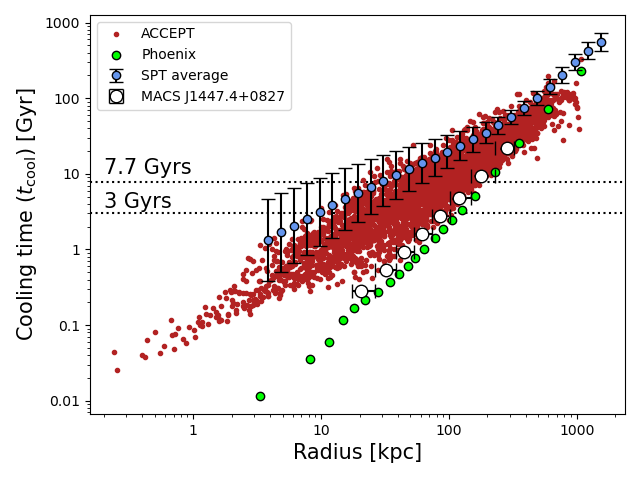}
\caption{\label{figtcool}{Average cooling time profile for the SPT clusters \citep{sanders_hydrostatic_2018} (blue circles) and cooling time profiles for 91 cool core clusters from the ACCEPT sample (red dots) and for the Phoenix cluster (green circles;  \citealt{mcdonald_anatomy_2019}), in which we overplotted the cooling time profile of MACS J447.4+0827 (white circles; vertical error bars are too small to be visible), extracted from the \textit{Chandra} deprojected thermodynamics quantities (for 6000 counts regions). Horizontal error bars indicate the radial extent of the bins used for fitting. We added dashed horizontal lines at $t_{\mathrm{cool}}$ = 3 Gyrs and 7.7 Gyrs for reference. Here, the cooling radius is defined as the radius where $t_{\mathrm{cool}}$ = 7.7 Gyrs, which equals to $\sim$ 133 kpc.}}
\end{figure}

Similar as the steps described in this section, we extracted the X-ray luminosity of the 9 regions using \texttt{XSPEC} in order to compute the cooling time profile. The model \texttt{clumin} -- that calculates X-ray luminosity between 0.01 and 50 keV -- was used and the data were deprojected using \texttt{projct}. The luminosity between 0.1 and 2.4 keV was also calculated (see Table \ref{tabxray} for the values). The cooling time arises from the following equation:

we\begin{equation}
t_{\mathrm{cool}}=\frac{5}{2}\frac{1.9 \, n_{\mathrm{e}} \, kT \, V}{L_{\mathrm{X}}},
\end{equation}

where $n_{\mathrm{e}}$ is the electronic density, $kT$ is the gas temperature, $V$ is the gas volume contained within an annulus and $L_{\mathrm{X}}$ is the gas X-ray luminosity. We define the cooling radius ($R_{\mathrm{cool}}$) as the radius for which the cooling time is equal to the $z = 1$ look-back time ($t_{\mathrm{cool}} = 7.7$ Gyrs; represented by a dashed green horizontal line in Fig. \ref{figtcool}), same as \cite{rafferty_feedback-regulated_2006}. We define the cooling luminosity ($L_{\mathrm{cool}}$) as the X-ray luminosity contained within $R_{\mathrm{cool}}$ between 0.1 and 2.4 keV. We extracted the X-ray luminosity within an ellipsoid with a semi-major axis of $a = R_{\mathrm{cool}}$ and an ellipticity of $\epsilon = 0.26$. This gives us $R_{\mathrm{cool}} \approx 133$ kpc and $L_{\mathrm{cool}} = (1.71 \pm 0.01) \times 10^{45}$ erg s$^{-1}$ (see Table \ref{tabxray}). We further analyze the implications of those results in Section \ref{sec:cavities}.

In Fig. \ref{figtcool}, we present the cooling time profile of 91 cool core clusters from the ACCEPT sample and of the Phoenix cluster \citep{mcdonald_anatomy_2019}, in which we overplotted MACS J1447.4+0827 and the average $t_\mathrm{cool}$ profile for the SPT clusters \citep{sanders_hydrostatic_2018}. Note that we kept the ACCEPT $t_\mathrm{cool}$ results, although they might not be accurate (e.g. \citealt{panagoulia_volume-limited_2014}), since they are frequently used in the literature and are here used as background purpose only. We come to the conclusion that MACS J1447.4+0827 is one of strongest cool core clusters known. The central point is influenced by the central AGN.

Furthermore, we added the model \texttt{mkcflow} based on \cite{mushotzky_einstein_1988} to calculate the mass deposition rate ($\dot{M}_{\mathrm{cool}}$) within the cooling radius $R_{\mathrm{cool}}$. We set the minimal temperature to 0.1 keV and the maximal temperature and abundance to the \texttt{apec} parameters. We obtained a 3$\sigma$ upper limit (see Table \ref{tabxray}).

\begin{table}[ht]
\centering
\caption{\label{tabxray}{X-ray properties of MACS J1447.4+0827.}}
	\begin{tabular}{cc}
	\hline\hline\\[-1.em]
	$\empty$ & Values\\
	\hline    
	$L_{\mathrm{X-ray} (0.01-50 \ \mathrm{keV})}$ & (5.31 $\pm$ 0.02) $\times$ 10$^{45}$ erg s$^{-1}$\\
    $L_{\mathrm{X-ray} (0.1-2.4 \ \mathrm{keV})}$ & (1.24 $\pm$ 0.01) $\times$ 10$^{45}$ erg s$^{-1}$\\
	$R_{\mathrm{cool}}$ & $\sim$ 133 kpc \\
	$L_{\mathrm{cool}}$ & (1.71 $\pm$ 0.01) $\times$ 10$^{45}$ erg s$^{-1}$ \\
	$\dot{M}_{\mathrm{cool}}$ & $<$ 96 M$_\odot$ yr$^{-1}$ \\
	\hline
	\end{tabular}
\end{table}

\subsection{Thermodynamic maps}\label{sec:maps}

In order to investigate the temperature and metal abundance distribution across the cluster, we generated thermodynamic maps using a technique mimicking that of \cite{kirkpatrick_hot_2015}. We first used a weighted Voronoi tessellation (WVT) algorithm \citep{cappellari_adaptive_2003,diehl_adaptive_2006} to bin the 0.5 -- 7 keV background-subtracted merged X-ray image. The WVT algorithm bins the image into similar signal-to-noise regions by first applying a bin accretion algorithm to the pixels in the image. The bin accretion map is then used as an initial guess for the true WVT algorithm which minimizes a predetermined scale length in order to more accurately group the pixels. The final result is a geometrically unbiased image of binned structures meeting a minimal signal-to-noise requirement. This technique has been implemented in \texttt{python v3.6} by us and can be found at \url{https://github.com/crhea93/AstronomyTools} under the Weighted Voronoi Tessellations repository.

With our pixel-binning scheme determined by the WVT algorithm, we must, for each individual observation, extract a spectrum corresponding to each bin. Once extracted for each bin, we simultaneously fit the three observations using \texttt{XSPEC v12.10.1}, \texttt{Sherpa v1}, and \texttt{python v3.5}. \texttt{Python} provided the main wrapping language, \texttt{Sherpa} enabled the use of well-developed fitting algorithms, and \texttt{XSPEC} supplied the requisite astrophysical models: \texttt{phabs} and \texttt{apec}. Constraining the redshift and column density, $n_{\mathrm{H}}$, to the previously mentioned values in Section \ref{sec:reduction}, we allow the temperature, metal abundance and normalization to be free. We employ the \texttt{c-statistic} determined by \texttt{Sherpa} in addition to the Nelder-Mead optimization algorithm. The entire temperature and abundance fitting algorithm can be found in its generalized form at \url{https://github.com/crhea93/AstronomyTools} under the Temperature Map Pipeline repository.

\begin{figure}[ht]
\includegraphics[width=\linewidth]{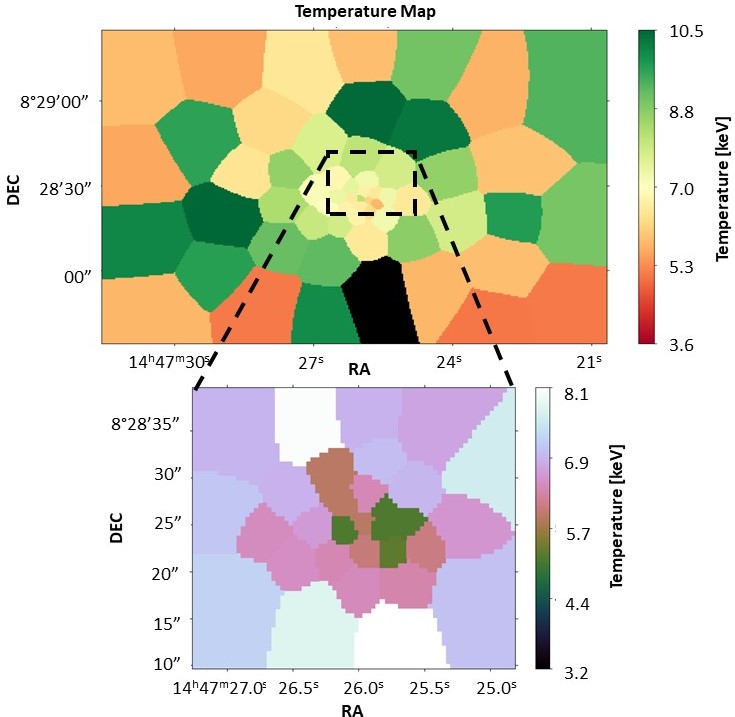}
\includegraphics[width=\linewidth]{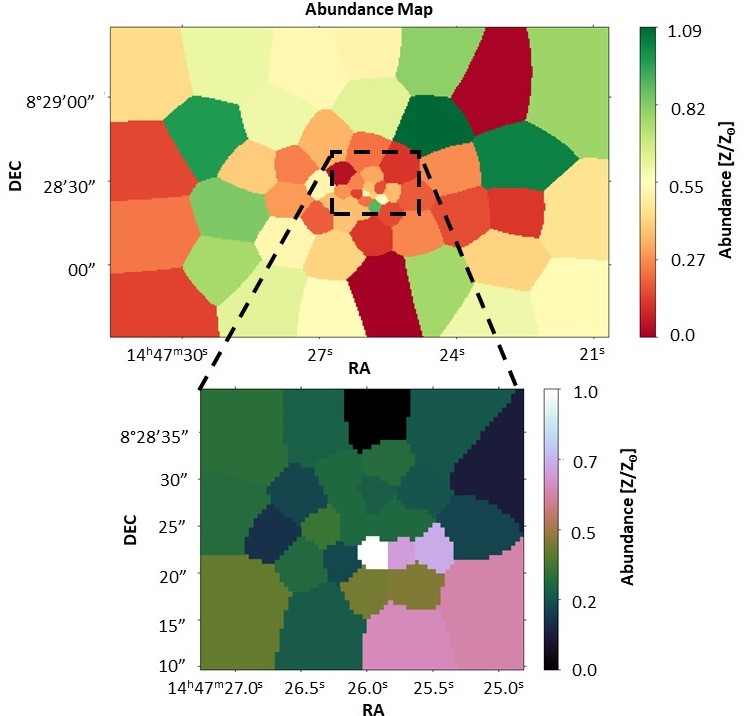}
\caption{\label{maps}{Large-scale temperature and abundance maps with a signal-to-noise ratio of 50. The insets are small-scale maps centered on the central regions with a signal-to-noise ratio of 40. The black regions signify a bin in which the fit failed to converge. Mean errors are 0.1 keV for the temperature and 0.05 Z$_\odot$ for the abundance.}}
\end{figure}

The large-scale (signal-to-noise ratio of 50) and small-scale (signal-to-noise ratio of 40) temperature and abundance maps are presented in Fig. \ref{maps}. We note that the temperature drops abruptly at the center by a factor of $\sim$ 3, which is consistent with a cool core. Those maps are also coherent with the thermodynamic profiles presented in Fig. \ref{figthermo}: we observe the same drop in temperature at the center and a similar abundance distribution.

\subsection{Characterization of the plume-like structure}\label{sec:plume}

As shown in the middle panel of Fig. \ref{fig:HARDSOFT}, the soft X-ray image reveals an extension of soft X-ray emission offset from the central AGN. It consists of a 2.8$\arcsec \times 3.1\arcsec$ plume-like structure centered at RA = $14^{\mathrm{h}}47^{\mathrm{m}} 25.825^{\mathrm{s}}$ and DEC = $+8^{\circ}28\arcmin 24.46\arcsec$ that encompasses some of the optical filaments of the \textsc{HST} image. We extracted the abundance and temperature of the plume using \texttt{phabs} and \texttt{apec}. We obtained T$_{\mathrm{plume}}$ = (4.8 $\pm$ 0.2) keV and Z$_{\mathrm{plume}}$ = (0.33 $\pm$ 0.06) Z$_\odot$. We also extracted a region surrounding the plume and obtained T$_{\mathrm{env}}$ = (5.8 $\pm$ 0.2) keV and Z$_{\mathrm{env}}$ = (0.30 $\pm$ 0.04) Z$_\odot$. We note that the plume is statistically colder than the surrounding medium. The colder region of the temperature map in Fig. \ref{maps} also corresponds with the plume-like structure.

A more detailed description of the plume-like structure and its implications will be discussed in the Section \ref{sec:ghostcav}.

\section{Radio analysis of the cluster}\label{sec:radiostruc}

In Fig. \ref{fig:radio_cont}, the \textsc{JVLA} A-configuration observations (left panel) reveal south-north oriented jetted outflows, where each lobe extends to $\sim$ 32 kpc in length, while B- and C-configuration observations (middle and right panels) reveal a diffuse radio structure extending to $\sim$ 160 kpc in size. In the following sections, we analyze these structures.

\subsection{Characterization of the relativistic jetted outflows}\label{sec:jets_chara}

We extracted the flux of the central radio AGN and its jetted outflows, using the 4$\sigma_{\mathrm{rms}}$ contours of the A-configuration observations and \texttt{CASA}. We extracted the flux using the tool \texttt{imstat} and estimated the error following \cite{cassano_revisiting_2013}:
\begin{equation}
    \sigma_{\mathrm{S}} = \sqrt{(\sigma_{\mathrm{cal}} S)^{2} + (\mathrm{rms}\sqrt{N_{\mathrm{beam}}})^{2} + \sigma_{\mathrm{sub}}^{2}},
    \label{eq:error}
\end{equation}

where $S$ is the flux, $\sigma_{\mathrm{cal}}$ is the percentage of the flux outside of the extraction region, rms is the image noise, $N_{\mathrm{beam}}$ is the number of beams in the region and $\sigma_{\mathrm{sub}}$ is the flux of point sources in the region. Here, we used $\sigma_{\mathrm{cal}} \sim 0\%$ (all the flux is contained in the extracted region), rms = 0.011 mJy/beam and $\sigma_{\mathrm{sub}} = 0$ (there are no point sources).

The radio power can be calculated using the following equation:

\begin{equation}
    P = 4 \pi S D_{\mathrm{L}}^2(1+z)^{-(\alpha +1)},
    \label{eq:power}
\end{equation}

where $D_{\mathrm{L}}$ is the luminosity distance and $\alpha$ is the spectral index, that characterizes the dependence of the spectra energy density $S_{\nu}$ in such a way that $S_{\mathrm{\nu}}$ $\propto$ $\nu^{\alpha}$. Assuming $\alpha = -1$ for jets, we obtained $P_{\mathrm{AGN+jets}}$ = (20.4 $\pm$ 0.1) $\times$ 10$^{24}$ W Hz$^{-1}$. We compare this result with the X-ray results in Section \ref{sec:cavities}.

\subsection{Characterization of the mini-halo}\label{sec:MH}

A radio mini-halo is a faint radio emission that has a characteristic size of 50 to 300 kpc found at the center of cool core clusters. \cite{richard} already presented an initial analysis of the radio mini-halo of MACS J1447.4+0827, but here, we focus on characterizing its radio properties compared to the X-ray properties of the cluster. 

The size and the flux of the mini-halo was calculated using the 4$\sigma_{\mathrm{rms}}$ contours of the C-configuration observations, since it contains 99$\%$ of the mini-halo's flux. The size can be found in Table \ref{tab:radioprop}. The tool \texttt{imstat} in \texttt{CASA} was used to extract the flux and we used $\sigma_{\mathrm{cal}} \sim 1\%$ (since 99\% of the mini-halo's flux is inside of the extraction region), rms = 0.015 mJy/beam and $\sigma_{\mathrm{sub}} = 0$ (there are no point sources) to estimate the error on the flux. However, the C-configuration contains the mini-halo's flux, but also the AGN's and its jets: we thus obtained the mini-halo's flux $S_{\mathrm{MH}}$ by subtracting the A-configuration flux ($S_{\mathrm{AGN+jets}}$) from the C-configuration flux ($S_{\mathrm{C}}$) (see Table \ref{tab:radioprop} for the values).

The mini-halo spectral index map of the configuration C at 1.4 GHz was calculated using the tool \texttt{clean}: it uses parameters that can model the frequency dependence across the bandwidth using the two first terms of the Taylor's series. The spectral map was obtained by dividing the image associated to the second term (\textit{image tt1}) by the image associated to the first term (\textit{image tt0}; total intensity of the 1.4 GHz image). Note that we only considered pixels detected at greater than 4$\sigma_{\mathrm{rms}}$. Taking the maximum value of an histogram of the spectral indexes of the map within a radius of $\sim$ $R_\mathrm{MH}$, we find $\alpha$ $\sim$ $-$1.2 for the mini-halo. The power of the mini-halo was calculated with the eq. \ref{eq:power} from the previous section and we obtained $P_{\mathrm{MH}} = (3.0 \pm 0.3)$ $\times$ 10$^{24}$ W Hz$^{-1}$. We further discuss these results in Section \ref{sec:disMH}. 

Table \ref{tab:radioprop} presents a summary of the radio properties of MACS J1447.4+0827.

\begin{table}[ht]
\centering
\caption{\label{tab:radioprop}{Radio properties in the L band (1.4 GHz).}} 
	\begin{tabular}{cc}
	\hline\hline\\[-1.em]
	$\empty$ & Values\\
	\hline
	$R_{\mathrm{jets}}$ & $\sim$ 32 kpc \\
	$R_{\mathrm{MH}}$ & $\sim$ 160 kpc\\
	$S_{\mathrm{AGN + jets}}$ & (39.4 $\pm$ 0.2) mJy \\
	$S_{\mathrm{MH}}$ & (5.7 $\pm$ 0.5) mJy \\
    $\alpha_{\mathrm{MH}}$ & $\sim$ $-$1.2\\
    $P_{\mathrm{AGN + jets}}$ & (19.1 $\pm$ 0.1) $\times$ 10$^{24}$ W Hz$^{-1}$\\
    $P_{\mathrm{MH}}$ & (3.0 $\pm$ 0.3) $\times$ 10$^{24}$ W Hz$^{-1}$\\
	\hline
	\end{tabular}
\end{table}

\section{The central AGN} \label{sec:agn}
\subsection{Radiative properties of the central AGN}\label{sec:rad}
Using the \textit{Chandra} 3 -- 8 keV image, we found evidence of a central point source whose position coincides with the brightest pixels of the \textsc{HST} image (RA = $14^{\mathrm{h}}47^{\mathrm{m}} 26.022^{\mathrm{s}}$, DEC = $+8^{\circ}28\arcmin 25.26\arcsec$) in the BCG: this emission is characteristic of an AGN. Taking a circular region with a radius of 1$\arcsec$ around the X-ray brightest pixel in the hard 3 -- 8 keV image, we compared the total number of counts to the 3$\sigma$ upper limit of the surrounding background. We defined the surrounding background as an annular region of 3$\arcsec$ of inner radius and of 5$\arcsec$ of outer radius. We calculated the number of counts, added 3$\sigma$ and scaled it to the number of counts expected for a small region of radius 1$\arcsec$. The total number of counts of the 1$\arcsec$ region is above the 3$\sigma$ upperlimit of the surrounding background, so the presence of an AGN is photometrically confirmed.

We furthermore calculated the hardness ratio defined as $\frac{H-S}{H+S}$, where $H$ is the number of counts in the hard X-rays and $S$ is the number of counts in the soft X-rays for a common region \citep{wang_identifying_2004,jin_hardness_2006,klesman_multi-wavelength_2012}. Taking the 0.5 -- 2 keV and 2 -- 8 keV images, we obtained a hardness ratio of = -0.25 $\pm$ 0.04 for the 1$\arcsec$ central region and a hardness ratio of -0.38 $\pm$ 0.01 for the 3$\arcsec$ $\times$ 5$\arcsec$ surrounding background. We note that the hardness ratio of the central region is 3$\sigma$ above the surrounding background. We therefore confirm the presence of an X-ray detected AGN in the BCG of MACS J1447.4+0827. 
We calculated the AGN's luminosity by first fitting the spectra of an annular 3 -- 5$\arcsec$ region to \texttt{phabs*apec}. We then extrapoled the temperature to the central 1$\arcsec$ region using the relation $T \propto r^{\mathrm{b}}$ (b $\sim$ 0.3; \citealt{voigt_thermal_2004}) and fitted a model \texttt{phabs*(powerlaw+apec)} to the region within r = 1$\arcsec$. Note that we fixed the abundance of the thermal gas and we fixed the photon index to $\Gamma$ = 1.9. We obtained an unabsorbed X-ray luminosity of $L_{\mathrm{AGN}}$ = 2.1 $\pm$ 0.2 $\times 10^{43}$ erg s$^{-1}$ between 2 -- 10 keV. We discuss the results in Section \ref{sec:poweringagn}.

\subsection{Radio spectral energy distribution}

\begin{figure}[ht]
    \centering
    \includegraphics[width=\linewidth]{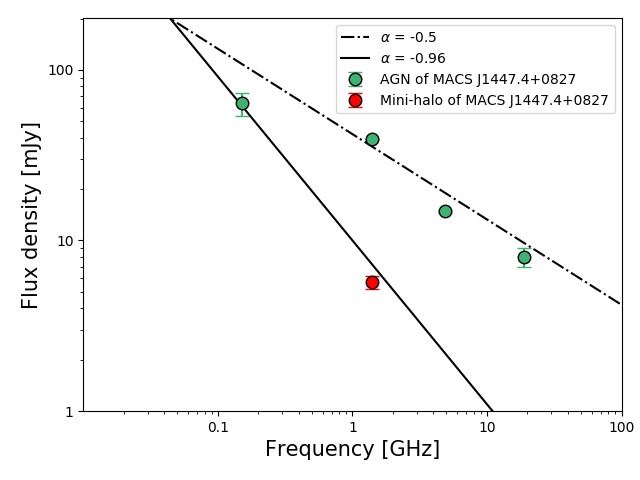}
    \caption{Radio spectral energy distribution (SED) of the BCG using data from TGSS, VLBA and ATCA for frequencies different from 1.4 GHz (private communication with A. C. Edge). The flux of the mini-halo (see Tab. \ref{tab:radioprop}) is also plotted. The dash-dotted black line indicates $\alpha = -0.5$ (fitting a power law using the 3 green dots on the right) and the plain black line indicates $\alpha = -0.96$ (fitting a power law using the green dot on the left and the red dot).}
    \label{fig:SED}
\end{figure}

In Fig. \ref{fig:SED}, we computed the spectral energy distribution (SED) of the cluster using the radio flux density measurements for the BCG in MACS J1447.4+0827 from our own analysis (see Section \ref{sec:jets_chara}) and from a private communication with Alastair C. Edge. We also plotted the mini-halo's flux (see Table \ref{tab:radioprop}). \cite{hogan_comprehensive_2015} found out that the radio SED can be decomposed into two major components: an active, typically flat spectrum component attributed to the AGN activity ($\alpha$ $>$ $-$0.5), and a steeper component that includes all of the others emissions ($\alpha < -0.5$).  We therefore plotted a line for $\alpha = -0.5$ for the 3 points on the right and a line for $\alpha = -0.96$ (average steeper spectrum emission) for the point on the left. The claimed mini-halo's flux in Table \ref{tab:radioprop} is consistent with the TGSS point. In fact, at low frequency, we see the past emission (or the jets). According to the flux equation, when the frequency decreases, the flux increases for old synchrotron emission. Consequently, the signal from the TGSS point comes from the past emission and not from the core. Also, MWA GLEAM upper limits are consistent with the $\alpha = -0.96$ line, but data from MWA or LOFAR could allow us to go deeper.

\subsection{Power of the X-ray cavities}\label{sec:power}
\begin{table*}
\centering
\caption{\label{tabcav}{X-ray cavity properties in MACS J1447.4+0827. The radii values indicate the minimum and maxium values from the X-ray images. Same for $P_{\mathrm{cav \, tot}}$.}}
	\begin{tabular}{ccccccc|c}
	\hline\hline \\[-1.em]
	Cavity & $R_{\mathrm{w}}$ & $R_{\mathrm{l}}$ & $R_{\mathrm{dist}}$ & $t_{\mathrm{sound}}$ & $P_{\mathrm{cav \, min}}$ & $P_{\mathrm{cav \, max}}$ & $P_{\mathrm{cav \, tot}}$\\
    $\empty$ & [kpc] & [kpc] & [kpc] & [10$^{7}$ yrs] & [10$^{44}$ erg s$^{-1}$] & [10$^{44}$ erg s$^{-1}$] & [10$^{44}$ erg s$^{-1}$]\\ 
	\hline
	Northern & [3.5; 4.2] & [5.0; 18.1] & 9.0 & 0.9 $\pm$ 0.1 & 2.5 $\pm$ 0.4 & 6.3 $\pm$ 1.1 & $P_{\mathrm{cav \, tot}}$ = [2.9 $\pm$ 0.4; 8.4 $\pm$ 1.1]\\
	Southern & [3.2; 3.9] & [4.1; 14.7] & 21.9 & 2.7 $\pm$ 0.2 & 0.40 $\pm$ 0.05 & 2.1 $\pm$ 0.3 & $P_{\mathrm{cav \, tot \, mean}}$ = 5.7 $\pm$ 0.5\\
	\hline
	\end{tabular}
\end{table*}

Here, we characterize the two X-ray cavities found in Fig. \ref{fig:subtracted} that coincide with the jetted outflows detected at radio wavelengths. If we assume that the cavities are in pressure equilibrium with the surrounding medium (e.g. \citealt{birzan_systematic_2004}), we can estimate the total energy contained in them with the following equation:
\begin{equation}
    E_{\mathrm{cavity}} = \frac{\gamma}{\gamma - 1}pV,
\end{equation}

where $\gamma$ is the heat capacity ratio, $p$ the thermal pressure and $V$ the volume of the cavity. For a relativistic fluid, $\gamma = 4/3$ \citep{nulsen_agn_2013} and therefore $E_{\mathrm{cavity}} = 4pV$.

We can approximate the volume of the cavity as $V = 4 \pi R_{\mathrm{w}}^2 R_{\mathrm{l}}/3$ assuming that they are ellipsoids. $R_{\mathrm{w}}$ and $R_{\mathrm{l}}$ are respectively defined as the projected semi-major axis perpendicular and parallel to the radio jet axis and have been measured directly based on Fig. \ref{fig:subtracted}. Note that we assign a 20 per cent uncertainty to $R_{\mathrm{w}}$ and $R_{\mathrm{l}}$ given the limited resolution of the X-ray data and because they are not deep enough to detect more clearly the cavities.

There are three different methods for computing the time scale associated with cavities: the buoyant rise time \citep{churazov_evolution_2001}, the refill time \citep{mcnamara_chandra_2000} and the sound crossing time defined as:

\begin{equation}
    t_{\mathrm{sound}} = \frac{R}{c_{\mathrm{s}}},
\end{equation}

where $c_{\mathrm{s}}$ the sound crossing time of the gas ($c_{\mathrm{s}} = \sqrt{\gamma kT/\mu m_{\mathrm{H}}}$). It remains unclear which one is the best estimation, but since they do not vary significantly from one another, we computed the cavity powers as $4pV/t_{\mathrm{sound}}$. Table \ref{tabcav} presents the results and we further discuss the consequences in Section \ref{sec:cavities}.







\section{Discussion} \label{sec:discussion}
\subsection{The strongest cool core clusters}

We have conducted an X-ray (\textit{Chandra}), radio (\textsc{JVLA}) and optical (\textsc{HST}) analysis of the massive cluster MACS J1447.4+0827. It is a strong cool core and X-ray luminous cluster. Its total mass, within a radius where the mean density is equal to 500 times the critical density at $z$ = 0.3755, is estimated to be $M_{500} = ($7.46$_{-0.86}^{+0.80}$) $\times$ 10$^{14}$ M$_\odot$ \citep{planck_collaboration_planck_2015}. 

\begin{figure}[ht]
\includegraphics[width=\linewidth]{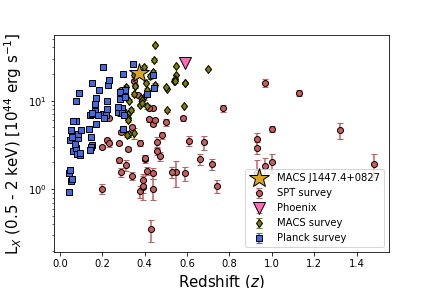}
\caption{\label{fig:complum}{X-ray luminosities in the 0.1 -- 2.4 keV energy range for a sample of clusters from the SPT (red circles; \citealt{bulbul_x-ray_2019}), MACS (green diamonds; \citealt{ebeling_complete_2007, ebeling_x-ray_2010}) and Planck (blue squares; \citealt{ade_planck_2011}) surveys. MACS J1447.4+0827 (yellow star) and Phoenix (pink triangle) are labelled. We note that MACS J1447.4+0827 is one of the X-ray brightest.}}
\end{figure}

In Fig. \ref{fig:complum}, we compare the 0.1 -- 2.4 keV X-ray luminosity of MACS J1447.4+0827 to other known clusters of galaxies with similar redshifts. This plot includes clusters from the \textit{MAssive Cluster Survey} (MACS; \citealt{ebeling_complete_2007, ebeling_x-ray_2010}), the \textit{South Pole Telescope} survey \citep{bulbul_x-ray_2019} and the \textit{Planck} survey \citep{ade_planck_2011}. MACS J1447.4+0827 is one of the most X-ray luminous clusters known in the literature.

\begin{figure}[ht]
\includegraphics[width=\linewidth]{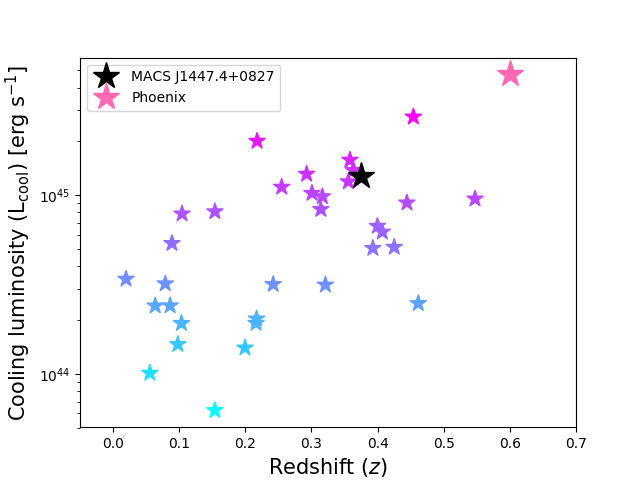}
\caption{\label{fig:comp}{Cooling luminosities in the 0.1 -- 2.4 keV energy range for a sample of strong cool core clusters from \cite{hlavacek-larrondo_probing_2013}. MACS J1447.4+0827 (black star) and Phoenix (pink star) are labelled.}}
\end{figure}

Furthermore, in Fig. \ref{fig:comp}, we compare MACS J1447.4+0827 to other strong cool core clusters. Here, we plot the cooling luminosity of several known cool core clusters as a function of redshift. Most of these data were taken from \cite{hlavacek-larrondo_rapid_2013}. While Fig. \ref{fig:comp} is not complete, we note that MACS J1447.4+0827 is one of the strongest cool core clusters around $z \approx 0.3$.

Several authors have also noted that strong cool core clusters often host a prominent abundance peak that coincides with the cool core (e.g. \citealt{ghizzardi_cold_2010,de_grandi_fe_2014}). Such peaks are thought to originate from metals being created by stellar processes in the central galaxy and have even been observed in high-redshift clusters (e.g. WARPS J1415.1+3612, \citealt{de_grandi_fe_2014}). However, as we can see on the up-right panel of Fig. \ref{figthermo}, the abundance profile of MACS J1447.4+0827 does not have a pronounced peak. The abundance profile shown in Fig. \ref{figthermo} further shows that the metallicity varies substantially region to region. This could be explained by recent sloshing motions that disturb the metal distribution close to the BCG. In fact, when the gas sloshes back and forth in the gravitational potential, it creates temperature and abundance discontinuities (e.g. \citealt{simionescu_metal_2010}). In Abell 3526 ($z$ = 0.01140), \cite{sanders_very_2016} showed that the sloshing motion can distort and induce asymmetries in the metal distribution across the BCG. Although MACS J1447.4+0827 has a much higher redshift than Abell 3526, we also observed a non-uniform metal distribution in the abundance profile in Fig. \ref{figthermo}. The metal distribution may trace the history of the gas motions and provide relevant information about the sloshing mechanisms (e.g. \citealt{simionescu_metal_2010}). 

\subsection{Relativistic outflows and X-ray cavities} \label{sec:cavities}
In \cite{hlavacek-larrondo_extreme_2012}, the authors analyzed \textit{Chandra} X-ray observations of 76 MACS clusters and found that 20 showed evidence of X-ray cavities. MACS J1447.4+0827 was found to host one potential X-ray cavity, although with the limited X-ray observations at the time ($\approx10$ ks), the authors were not able to study the properties of the cavity in detail. 

Here, using new deep \textit{Chandra} observations, as well as new \textsc{JVLA} observations, we find clear evidence of two collimated jetted outflows, one of which coincides with the X-ray cavity detected by \cite{hlavacek-larrondo_extreme_2012}. The two X-ray cavities are located symmetrically to the north and south of the central galaxy. They both have a radius of $\approx16$ kpc, although they have an elliptical and not circular morphology. Interestingly, the jetted outflows are clearly detected at radio wavelength, but they are only detected at a $\sim$ 2$\sigma$ level compared to the surrounding X-ray background, which is not the case for other known cavities at similar redshifts. 

In \cite{kolokythas_complete_2018}, the authors compared the X-ray cavity power $P_{\mathrm{cav}}$ to the radio power of the central AGN at 235 MHz and found that the two quantities are correlated to each other. In MACS J1447.4+0827, we found that $P_{\mathrm{cav}} \sim 6 \times 10^{44}$ erg s$^{-1}$ and $P_{\mathrm{AGN+jets}}$ = (20.4 $\pm$ 0.1) $\times$ 10$^{24}$ W Hz$^{-1}$. By converting our radio power at 1.4 GHz to 235 MHz assuming $\alpha \approx -1$, we find that MACS J1447.4+0827 falls directly on the relations found by \cite{birzan_radiative_2010} and \cite{osullivan_heating_2011} (see Fig. 7 of \cite{kolokythas_complete_2018}), implying that the radio powers and X-ray cavity powers in this cluster are consistent with those detected in other systems. 

\begin{figure}[ht]
    \centering
    \includegraphics[width=\linewidth]{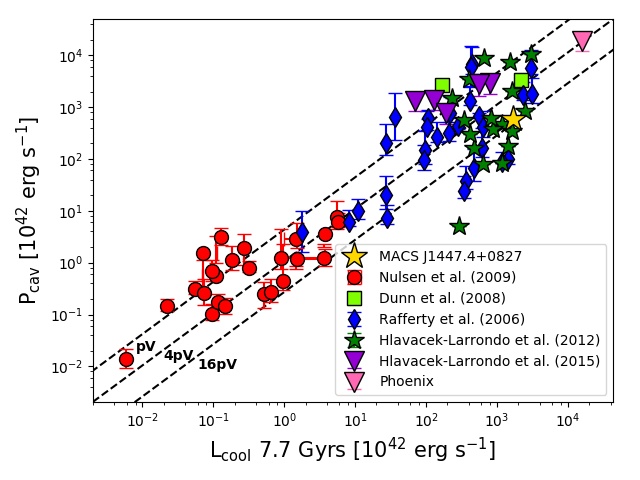}
    \caption{Total mechanical power ($P_{\mathrm{cav}}$) of the X-ray cavities as a function of the cooling luminosity ($L_{\mathrm{cool}}$) at 7.7 Gyrs. Data from \cite{rafferty_feedback-regulated_2006} is indicated as blue diamonds, data from \cite{dunn_investigating_2008} as green squares, data from \cite{nulsen_radio_2009} as red circles, data from \cite{hlavacek-larrondo_hunt_2012} as green stars and data from \cite{hlavacek-larrondo_x-ray_2015} as purple triangles. MACS J1447.4+0827 (yellow star) and Phoenix (pink triangle; \citealt{mcdonald_anatomy_2019}) are labelled. Dashed black lines denote $P_{\mathrm{cav}}$ = $L_{\mathrm{cool}}$ for $E_{\mathrm{cav}}$ = $pV, 4pV$ and $16pV$. Figure adapted from \cite{hlavacek-larrondo_x-ray_2015}.}
    \label{fig:dereg}
\end{figure}

In Fig. \ref{fig:dereg}, we compare the total mechanical power of X-ray cavities to the cooling luminosity of various samples of clusters. Here, the figure has been adapted from \cite{hlavacek-larrondo_x-ray_2015} and the data taken from \cite{rafferty_feedback-regulated_2006}, \cite{dunn_investigating_2008}, \cite{nulsen_radio_2009} and \cite{hlavacek-larrondo_hunt_2012,hlavacek-larrondo_x-ray_2015}. 

The total mechanical power of the X-ray cavities in MACS J1447.4+0827 seems to be lower than the cooling luminosity by a factor of $\sim$ 8. As we see on the Fig. \ref{fig:dereg}, MACS J1447.4+0827 appears to be closer to the $E_{\mathrm{cav}}$ = $16pV$ line than the $E_{\mathrm{cav}}$ = $4pV$ line. This could mean that MACS J1447.4+0827 is in a phase of lower AGN power: depending on the different cycles of AGN feedback, the power of the X-ray cavities can vary, although on average the power is sufficient to suppress cooling of the hot atmosphere.  However, considering the (large) uncertainties in X-ray cavity powers, the scatter in Fig. \ref{fig:dereg} and the fact that MACS J1447.4+0827 falls within this scatter, AGN feedback can easily be strong enough to offset cooling in this cluster. AGN feedback therefore seems to be, at the very least, present in the strongest cool core clusters at $z$ = 0.3 and may still provide a viable solution to the cooling flow problem even at these redshifts.

\begin{figure}[ht]
\includegraphics[width=\linewidth]{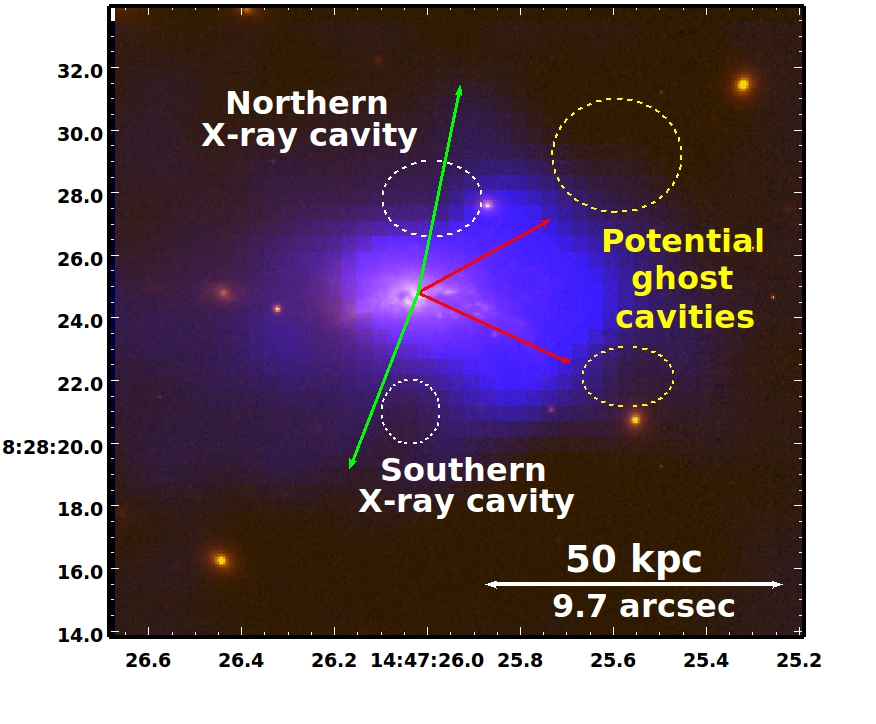}
\caption{\label{fig:fleches}{Combined \textsc{HST} F814W (red), F606W (green) and double-$\beta$ subtracted \textit{Chandra} (blue) images of MACS J1447.4+0827. The green arrows show the direction of jetted outflows based on the \textsc{JVLA} data, and the red arrows show the prominent filaments based on the \textsc{HST} data. The approximate location of the northern and southern cavities are indicated, as well as the approximate location of the potential ghost cavities based on Fig. \ref{fig:subtracted}.}}
\end{figure}

On the right panel of Fig. \ref{fig:subtracted}, we also note that the direction of the outflows matches the location of the northern cavity, but not exactly the southern cavity. To highlight this better, we present the combined \textsc{HST} F814W (red), F606W (green) and double-$\beta$ subtracted \textit{Chandra} (blue) images of MACS J1447.4+0827 in Fig. \ref{fig:fleches}, in which we indicated the direction of the radio outflows (green arrows) and prominent optical filaments (red arrows), as well as the approximate location of the X-ray cavities (white ellipses) and ghost cavities (yellow ellipses) as seen in the unsharp-masked and double-$\beta$ subtracted \textit{Chandra} images. Note that the filaments and ghost cavities are discussed in the subsequent section. The shift between the south-oriented outflow and the southern cavity is interesting and may be explained by the dynamic nature of the cluster, although deeper observations are needed to better understand this scenario.

There are a number of lines of evidence for the presence of sloshing motions in MACS J1447.4+0827. The temperature profile (Fig. \ref{figthermo}),  the plume-like structure (Fig. \ref{fig:HARDSOFT}), the optical filaments (Fig. \ref{fighubble}) and the prominent ellipticity of the cluster all point to a past merger event. However, as we do not detect another X-ray peak, we suspect that it was a relatively minor merger event. This merger event and the sloshing it generates could therefore induce a shift between the cool core and the BCG \citep{hamer_relation_2012}.


\subsection{Optical filaments, ghost cavities and the plume}\label{sec:ghostcav}

In Section \ref{sec:tcool}, we introduced the possibility of the presence of ghost cavities, consisting of cavities that have been dragged out of the AGN's area to larger radii. This type of cavity have been seen in other clusters at lower redshifts (e.g. Abell 2052, \citealt{blanton_shocks_2009}; Abell 2597, \citealt{mcnamara_discovery_2001}; Fornax, \citealt{su_buoyant_2017}; Abell 1795, \citealt{walker_exploring_2014}; Perseus, \citealt{hitomi_collaboration_atmospheric_2018}), but also at higher redshifts (e.g. Pheonix, \citealt{mcdonald_deep_2015}; RX J1532.9+3021, \citealt{hlavacek-larrondo_probing_2013}).

In Fig. \ref{fig:fleches}, we note that the prominent optical filaments of the cluster are oriented towards the potential ghost cavities. Optical filaments are thought to be created from thermally unstable ICM gas (e.g. \citealt{mccourt_thermal_2012,sharma_thermal_2012,gaspari_chaotic_2013,li_modeling_2014,li_modeling_2014-1}) that originates, in part, from the pushing-back motions created by the formation of X-ray cavities (e.g. \citealt{mcnamara_mechanism_2016,hogan_onset_2017,prasad_agn_2017,voit_global_2017,gaspari_shaken_2018}). Therefore, filaments could have been created in the wake of the cavities (e.g. \citealt{hlavacek-larrondo_probing_2013}), and thus the observed optical filaments in MACS J1447.4+0827 could be created in the wake of the ghost cavities seen to the west. These X-ray cavities could have been pushed back by sloshing motions towards the western side of the cluster, implying that X-ray cavities do not just rise buoyantly, but that there are also constantly affected by the dynamics of the X-ray gas in the cluster. Due to the evidence of sloshing motions mentioned before, the bulk of the ICM may have moved to the west, which may have caused a shift between the jets and part of the southern X-ray cavity. Such dynamic motions of the cluster may also have shifted the position of the ghost cavities in the same direction. 

In Section \ref{sec:plume}, we characterized the X-ray plume-like structure. We found out that the plume was colder than the environment. The origin of this type of structure is still unsure, but has been observed in many clusters (e.g. Abell 3526, \citealt{sanders_spatially_2002,crawford_extended_2005,walker_x-ray_2015}; RX J1532.9+3021, \citealt{hlavacek-larrondo_probing_2013}; merging Abell 2146, \citealt{russell_shock_2012}). The plume would come from the gas from the regions in the BCG: the gas would have been first driven out by the cavities, before falling back to the BCG (e.g. \citealt{walker_x-ray_2015}). In their study of the optical filaments in NCG 1275, \cite{fabian_energy_2011} proposed that the gas in the plume is excited by the ICM gas. While cooling, the hot gas from the ICM penetrates the optical cold filaments through the reconnection diffusion process which allows the hot gas to enter the cold gas despite the magnetic field that holds the filaments. By crossing it, the gas of the ICM transfers energy to the filaments and would allow its excitation. 

\subsection{Radio mini-halo}\label{sec:disMH}

There are now over 30 radio mini-halos that have been discovered (see \citealt{giacintucci_expanding_2019} and \citealt{richard}). The mini-halo of MACS J1447.4+0827 was presented and analyzed in detail in \cite{richard}. The mini-halo's power was plotted as a function of the 0.01 -- 50 keV X-ray luminosity and compared to other mini-halos in cool core clusters. It was found out that the mini-halo in MACS J1447.4+0827 follows this correlation but is one of the most radio powerful known to date. 

The exact origin of mini-halos is still unknown. Like jetted outflows, their radiation is due to synchrotron emission by relativistic particles in the cluster's magnetic field. However, the radiative lifetime of the particles is too short to spread over the observed mini-halo's radius. This implies that the particles of mini-halos come from the re-acceleration of relativistic particles or are produced in-situ \citep{van_weeren_diffuse_2019}. For the latter, mini-halos could come from the proton-proton interaction between the cosmic rays and the ICM \citep{pfrommer_constraining_2004,fujita_nonthermal_2007,zandanel_physics_2014}. For the former, it has been proposed that turbulence in the ICM could be the source of re-acceleration. It was suggested that this turbulence could either be caused by the dynamics of the jets (e.g. \citealt{bravi_connection_2016, richard}) or by sloshing motions of the ICM via minor mergers \citep{mazzotta_radio_2008,zuhone_turbulence_2013,giacintucci_mapping_2014}. This kind of motion leaves traces: cold fronts, which are discontinuities of temperature and density that appear to bound the mini-halos (e.g. \citealt{markevitch_non-hydrostatic_2001,mazzotta_chandra_2001,mazzotta_chandra_2003,markevitch_shocks_2007,owers_high_2009,ghizzardi_cold_2010}). 

In MACS J1447.4+0827, we find evidence of 2 cold fronts (see up-left panel of Fig. \ref{figthermo}) located at a radius of $\sim$ 50 kpc and $\sim$ 300 kpc. The mini-halo in MACS J1447.4+0827 therefore appears to be bounded by the cold fronts, since $R_{\mathrm{MH}}$ $\sim$ 160 kpc. However, \cite{richard} recently found evidence that mini-halos are directly connected to the AGN feedback processes of the BCG. In this work, the power of the mini-halo was plotted as a function of the BCG steep radio power at 1 GHz, the BCG core radio power at 10 GHz and the total power of the X-ray cavities. In all cases, the mini-halo of MACS J1447.4+0827 agrees with the correlations, indicating that the mini-halo in this cluster could also be connected at a fundamental level with its feedback processes. 

\subsection{Power mechanisms in the central AGN} \label{sec:poweringagn}

In order to further understand the feedback mechanisms occurring in the central AGN of MACS J1447.4+0827, we computed its Eddington luminosity, the black hole's mean accretion rate and the Eddington rate. 

The Eddington luminosity is defined as the limiting luminosity where the radiation pressure is in equilibrium with the local gravity of the accreting material. For a fully ionized plasma, the Eddington luminosity is given by the following equation:

\begin{equation}
    \frac{L_{\mathrm{Edd}}}{\mathrm{erg} \, \mathrm{s}^{-1}} = 1.26 \times 10^{47} \Bigg(\frac{M_{\mathrm{BH}}}{10^9 M_{\odot}}\Bigg),
\end{equation}

where $M_{\mathrm{BH}}$ is the SMBH mass. Since there are no mass estimates of the black hole available in the literature, we assumed a mass of between 9 and 10 $\times$ 10$^{9}$ M$_{\odot}$ for MACS J1447.4+0827, which is the typical mass expected for such a massive cluster and central galaxy.

The mean accretion rate scaled by the Eddington rate can be expressed as:

\begin{equation}
   \frac{\dot{M}}{\dot{M}_{\mathrm{Edd}}} = \frac{(P_{\mathrm{cav}} + L_{\mathrm{bol}})}{L_{\mathrm{Edd}}},
\end{equation}

where $L_{\mathrm{bol}}$ is the bolometric luminosity. Here, we used a typical bolometric correction of 5 for low luminosity sources on the 2 -- 10 keV X-ray luminosity of the central AGN (see Section \ref{sec:rad} for the value), following in \cite{russell_radiative_2013}.

\begin{figure}[ht]
    \centering
    \includegraphics[width=\linewidth]{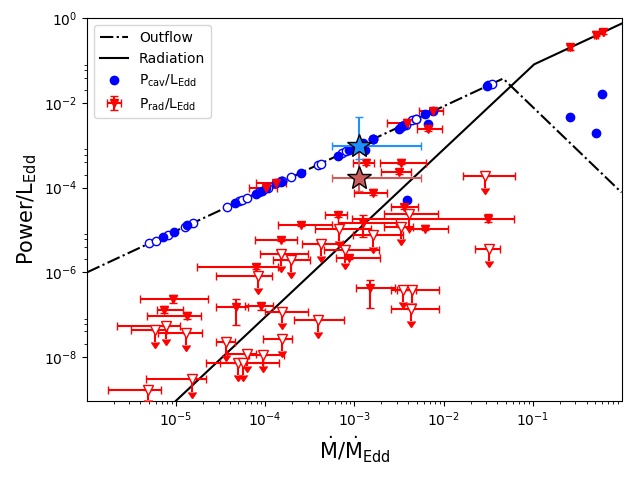}
    \caption{The cavity power (blue circles) and the radiative power (red triangles) scaled by the Eddington luminosity as a function of the required mean accretion rate scaled by the Eddington rate ($\dot{M}/\dot{M}_{\mathrm{Edd}}$). The open symbols indicates the upperlimits, which are on the X-ray luminosity. Figure adapted from \cite{russell_radiative_2013}. There are two points for each source. MACS J1447.4+0827 is labelled as star symbols.}
    \label{fig:edd}
\end{figure}

Fig. \ref{fig:edd} presents the cavity power (blue circles) and the radiative power (red triangles) scaled by the Eddington luminosity as a function of the required mean accretion rate scaled by the Eddington rate ($\dot{M}/\dot{M}_{\mathrm{Edd}}$). The open symbols indicates the upperlimits. The figure was adapted from \cite{russell_radiative_2013} (Fig. 12) and \cite{churazov_supermassive_2005} (Fig. 1). We plotted the location of MACS J1447.4+0827 (see star symbols), similarly with the cavity power in blue and radiative power and red. 

At low accretion rates, it is the jetted outflow mode that is dominant rather than the radiative (e.g. \citealt{churazov_supermassive_2005,russell_radiative_2013}). Indeed, Fig. \ref{fig:edd} shows that for $\dot{M}/\dot{M}_{\mathrm{Edd}}$ $\lesssim$ 3 $\times$ 10$^{-2}$, the outflow mode is dominating over the radiative mode. Since MACS J1447.4+0827 is not in the quasar phase, the outflow mode is dominating over the radiative mode by a factor of $\sim$ 10.

The radiative mode would act mainly in the young AGNs ($z$ $\gtrsim$ 1) when they are in the quasar stage and the accretion rate of the black hole is high (e.g. \citealt{churazov_supermassive_2005,hopkins_unified_2006}). We find that our object is perfectly consistent with the low-accretion mode, even though it is located at $z>0.3$. We can therefore affirm that this feedback mode has been in place several Giga-years ago in MACS J1447.4+0827. It is also happening in many other clusters (e.g. \citealt{hlavacek-larrondo_hunt_2012, hlavacek-larrondo_x-ray_2015}).

\section{Summary and Conclusions}

In this work, we have presented the first detailed and multiwavelength analysis of one of the strongest cool core cluster known, MACS J1447.4+0827 ($z$ = 0.3755) using data from the \textit{Hubble Space Telescope}, and new data from the \textit{Chandra X-ray Observatory} and the Karl G. Janksy Very Large Array. Our relevant findings are summarized as follows:

\begin{itemize}
    \item MACS J1447.4+0827, with its 0.01 -- 50 keV X-ray luminosity of $L_{\mathrm{X}}$ = (5.31 $\pm$ 0.02) $\times$ 10$^{45}$ erg s$^{-1}$, is one of the most luminous clusters at $z$ $<$ 0.7.
	\item We detected a northern and southern X-ray cavity as well as two potential ghost cavities. The A-configuration \textsc{JVLA} observations reveals south-north oriented jetted outflows extending to $\sim$ 32 kpc in length and coincident with the X-ray cavities. Their total mechanical power is $P_{\mathrm{cav}}$ $\approx 6 \times 10^{44}$ erg s$^{-1}$.
    \item We calculated the cooling luminosity of $L_{\mathrm{cool}} = (1.71 \pm 0.01) \times 10^{45}$ erg s$^{-1}$. This cooling luminosity indicates that MACS J1447.4+0827 is one of the strongest cool core clusters around $z$ $<$ 0.7. We found that the total mechanical power of the X-ray cavities was generally consistent with power required to offset the cooling of the ICM. MACS J1447.4+0827 might currently be going through a phase of lower activity, but not to the point to lead to a catastrophic cooling of the core.
    \item The \textsc{HST} images reveal an
    elliptic BCG in the centre of MACS J1447.4+0827 as well as large H$\alpha$ filaments whose directions match the location of the potential ghost cavities. Part of these filaments also match the position of a cool X-ray plume.
    \item The B- and C-configuration \textsc{JVLA} observations reveal a diffuse radio structure extending to $\sim$ 160 kpc in size interpreted as a mini-halo. The total radio power is $P_{1.4 \mathrm{GHz}} = (3.0 \pm 0.3) \times 10^{24}$ W Hz$^{-1}$. It is one of the most powerful mini-halos found.
\end{itemize}

    
While this study clearly shows that powerful mechanical feedback is occurring in the strongest cool core clusters, new and/or deeper observations in other wavelengths such as submillimiter, infrared, optical and ultraviolet -- with the Multi Unit Spectroscopic Explorer (MUSE) -- could provide more information concerning the star formation rate, the quantity and location of molecular gas in the BCG and would allow to better understand the multiphase gas in extreme cool core clusters. 

\section*{Acknowledgments}
We wish to thank M. McDonald for sharing his data and for his help.

MPE is supported by NRSEC (Natural Sciences and Engineering Research Council of Canada) and by FRQNT (Fonds de recherche du Qu\'{e}bec - Nature et technologies) through the Undergraduate Student Research Awards, and by IVADO (Institute for Data Valorization) through the Summer Internship and Undergraduate Research Fellowships. JHL is supported by NSERC through the Canada Research Chair programs and wish to acknowledge the support of an NSERC Discovery Grant and of the FRQNT. ARL is supported by NSERC through the NSERC Alexander-Graham-Bell Canada Graduate Scholarships-Master's Program (CGS M) and by FRQNT through the FRQNT Graduate Studies Research Scholarship - M. Sc. level under grant \#209839.

\bibliographystyle{aasjournal}
\bibliography{lib}

\begin{thebibliography}{}
\expandafter\ifx\csname natexlab\endcsname\relax\def\natexlab#1{#1}\fi
\providecommand{\url}[1]{\href{#1}{#1}}

\bibitem[{Anders \& Grevesse(1989)}]{anders_abundances_1989}
Anders, E., \& Grevesse, N. 1989, Geochimica et Cosmochimica Acta, 53, 197

\bibitem[{Applegate {et~al.}(2014)Applegate, von~der Linden, Kelly, Allen,
  Allen, Burchat, Burke, Ebeling, Mantz, \& Morris}]{applegate_weighing_2014}
Applegate, D.~E., von~der Linden, A., Kelly, P.~L., {et~al.} 2014, MNRAS, 439,
  48

\bibitem[{Applegate {et~al.}(2016)Applegate, Mantz, Allen, von~der Linden,
  Morris, Hilbert, Kelly, Burke, Ebeling, Rapetti, \&
  Schmidt}]{applegate_cosmology_2016}
Applegate, D.~E., Mantz, A., Allen, S.~W., {et~al.} 2016, MNRAS, 457, 1522

\bibitem[{Birzan {et~al.}(2008)Birzan, {McNamara}, Nulsen, Carilli, \&
  Wise}]{birzan_radiative_2010}
Birzan, L., {McNamara}, B.~R., Nulsen, P. E.~J., Carilli, C.~L., \& Wise, M.~W.
  2008, {ApJ}, 709, 546

\bibitem[{Birzan {et~al.}(2004)Birzan, Rafferty, {McNamara}, Wise, \&
  Nulsen}]{birzan_systematic_2004}
Birzan, L., Rafferty, D.~A., {McNamara}, B.~R., Wise, M.~W., \& Nulsen, P.
  E.~J. 2004, {ApJ}, 607, 800

\bibitem[{Blanton {et~al.}(2009)Blanton, Randall, Douglass, Sarazin, Clarke, \&
  {McNamara}}]{blanton_shocks_2009}
Blanton, E.~L., Randall, S.~W., Douglass, E.~M., {et~al.} 2009, ApJ Letters,
  697, L95

\bibitem[{Bravi {et~al.}(2016)Bravi, Gitti, \&
  Brunetti}]{bravi_connection_2016}
Bravi, L., Gitti, M., \& Brunetti, G. 2016, arXiv e-prints, arXiv:1603.00368

\bibitem[{Bulbul {et~al.}(2019)Bulbul, Chiu, Mohr, {McDonald}, Benson, Bautz,
  Bayliss, Bleem, Brodwin, Bocquet, Capasso, Dietrich, Forman,
  Hlavacek-Larrondo, Holzapfel, Khullar, Klein, Kraft, Miller, Reichardt, Saro,
  Sharon, Stalder, Schrabback, \& Stanford}]{bulbul_x-ray_2019}
Bulbul, E., Chiu, I.-N., Mohr, J.~J., {et~al.} 2019, {ApJ}, 871, 50

\bibitem[{Cappellari \& Copin(2003)}]{cappellari_adaptive_2003}
Cappellari, M., \& Copin, Y. 2003, MNRAS, 342, 345

\bibitem[{Cassano {et~al.}(2013)Cassano, Ettori, Brunetti, Giacintucci, Pratt,
  Venturi, Kale, Dolag, \& Markevitch}]{cassano_revisiting_2013}
Cassano, R., Ettori, S., Brunetti, G., {et~al.} 2013, {ApJ}, 777, 141

\bibitem[{Castillo-Morales \& Schindler(2003)}]{castillo-morales_clusters_2003}
Castillo-Morales, A., \& Schindler, S. 2003, {arXiv}:astro-ph/0303610,
  astro-ph/0303610

\bibitem[{Cavagnolo {et~al.}(2008)Cavagnolo, Donahue, Voit, \&
  Sun}]{cavagnolo_entropy_2008}
Cavagnolo, K.~W., Donahue, M., Voit, G.~M., \& Sun, M. 2008, {ApJ}, 683, L107

\bibitem[{Cavagnolo {et~al.}(2010)Cavagnolo, {McNamara}, Nulsen, Carilli,
  Jones, \& Birzan}]{cavagnolo_relationship_2010}
Cavagnolo, K.~W., {McNamara}, B.~R., Nulsen, P. E.~J., {et~al.} 2010, {ApJ},
  720, 1066

\bibitem[{Cavaliere \& Fusco-Femiano(1976)}]{cavaliere_x-rays_1976}
Cavaliere, A., \& Fusco-Femiano, R. 1976, A \& A, 49, 137

\bibitem[{Chirivì {et~al.}(2018)Chirivì, Suyu, Grillo, Halkola, Balestra,
  Caminha, Mercurio, \& Rosati}]{chirivi_macs_2018}
Chirivì, G., Suyu, S.~H., Grillo, C., {et~al.} 2018, A \& A, 614, A8

\bibitem[{Churazov {et~al.}(2001)Churazov, Brüggen, Kaiser, Böhringer, \&
  Forman}]{churazov_evolution_2001}
Churazov, E., Brüggen, M., Kaiser, C.~R., Böhringer, H., \& Forman, W. 2001,
  {ApJ}, 554, 261

\bibitem[{Churazov {et~al.}(2005)Churazov, Sazonov, Sunyaev, Forman, Jones, \&
  Boehringer}]{churazov_supermassive_2005}
Churazov, E., Sazonov, S., Sunyaev, R., {et~al.} 2005, MNRAS: Letters, 363, L91

\bibitem[{Crawford {et~al.}(2005)Crawford, Hatch, Fabian, \&
  Sanders}]{crawford_extended_2005}
Crawford, C.~S., Hatch, N.~A., Fabian, A.~C., \& Sanders, J.~S. 2005, MNRAS,
  363, 216

\bibitem[{De~Filippis {et~al.}(2002)De~Filippis, Schindler, \&
  Castillo-Morales}]{de_filippis_recent_2002}
De~Filippis, E., Schindler, S., \& Castillo-Morales, A. 2002,
  {arXiv}:astro-ph/0201349, astro-ph/0201349

\bibitem[{De~Grandi {et~al.}(2014)De~Grandi, Santos, Nonino, Molendi, Tozzi,
  Rossetti, Fritz, \& Rosati}]{de_grandi_fe_2014}
De~Grandi, S., Santos, J.~S., Nonino, M., {et~al.} 2014, A\&A, 567, A102

\bibitem[{Diehl \& Statler(2006)}]{diehl_adaptive_2006}
Diehl, S., \& Statler, T.~S. 2006, MNRAS, 368, 497

\bibitem[{Dong {et~al.}(2010)Dong, Ho, Wang, Wang, Wang, Fan, \&
  Zhou}]{dong_prevalence_2010}
Dong, X.-B., Ho, L.~C., Wang, J.-G., {et~al.} 2010, {ApJ}, 721, L143

\bibitem[{Dunn \& Fabian(2006)}]{dunn_investigating_2006}
Dunn, R. J.~H., \& Fabian, A.~C. 2006, MNRAS, 373, 959

\bibitem[{Dunn \& Fabian(2008)}]{dunn_investigating_2008}
---. 2008, Monthly Notices {RAS}, 385, 757

\bibitem[{Ebeling {et~al.}(2007)Ebeling, Barrett, Donovan, Ma, Edge, \& van
  Speybroeck}]{ebeling_complete_2007}
Ebeling, H., Barrett, E., Donovan, D., {et~al.} 2007, ApJ, 661, L33

\bibitem[{Ebeling {et~al.}(2001)Ebeling, Edge, \& Henry}]{ebeling_macs:_2001}
Ebeling, H., Edge, A.~C., \& Henry, J.~P. 2001, ApJ, 553, 668

\bibitem[{Ebeling {et~al.}(2010)Ebeling, Edge, Mantz, Barrett, Henry, Ma, \&
  van Speybroeck}]{ebeling_x-ray_2010}
Ebeling, H., Edge, A.~C., Mantz, A., {et~al.} 2010, MNRAS, 407, 83

\bibitem[{Ehlert {et~al.}(2011)Ehlert, Allen, von~der Linden, Simionescu,
  Werner, Taylor, Gentile, Ebeling, Allen, Applegate, Dunn, Fabian, Kelly,
  Million, Morris, Sanders, \& Schmidt}]{ehlert_extreme_2011}
Ehlert, S., Allen, S., von~der Linden, A., {et~al.} 2011, MNRAS, 411, 1641

\bibitem[{Fabian(1999)}]{fabian_massive_1999}
Fabian, A. 1999, Chandra Proposal

\bibitem[{Fabian(1994)}]{fabian_cooling_1994}
Fabian, A.~C. 1994, Annu. Rev. Astron. Astrophys., 32, 277

\bibitem[{Fabian {et~al.}(2011)Fabian, Sanders, Williams, Lazarian, Ferland, \&
  Johnstone}]{fabian_energy_2011}
Fabian, A.~C., Sanders, J.~S., Williams, R. J.~R., {et~al.} 2011, MNRAS, 417,
  172

\bibitem[{Farrah {et~al.}(2016)Farrah, Baloković, Stern, Harris, Kunimoto,
  Walton, Alexander, Arévalo, Ballantyne, Bauer, Boggs, Brandt, Brightman,
  Christensen, Clements, Craig, Fabian, Hailey, Harrison, Koss, Lansbury, Luo,
  Paine, Petty, Pitchford, Ricci, \& Zhang}]{farrah_geometry_2016}
Farrah, D., Baloković, M., Stern, D., {et~al.} 2016, ApJ, 831, 76

\bibitem[{Franceschini {et~al.}(2000)Franceschini, Bassani, Cappi, Granato,
  Malaguti, Palazzi, \& Persic}]{franceschini_bepposax_2000}
Franceschini, A., Bassani, L., Cappi, M., {et~al.} 2000, A \& A, 353, 910

\bibitem[{Fujita {et~al.}(2007)Fujita, Kohri, Yamazaki, \&
  Kino}]{fujita_nonthermal_2007}
Fujita, Y., Kohri, K., Yamazaki, R., \& Kino, M. 2007, ApJ Letters, 663, L61

\bibitem[{Gaspari {et~al.}(2013{\natexlab{a}})Gaspari, Brighenti, \&
  Ruszkowski}]{gaspari_solving_2013}
Gaspari, M., Brighenti, F., \& Ruszkowski, M. 2013{\natexlab{a}}, Astron.
  Nachr., 334, 394

\bibitem[{Gaspari {et~al.}(2013{\natexlab{b}})Gaspari, Ruszkowski, \&
  Oh}]{gaspari_chaotic_2013}
Gaspari, M., Ruszkowski, M., \& Oh, S.~P. 2013{\natexlab{b}}, MNRAS, 432, 3401

\bibitem[{Gaspari {et~al.}(2018)Gaspari, {McDonald}, Hamer, Brighenti, Temi,
  Gendron-Marsolais, Hlavacek-Larrondo, Edge, Werner, Tozzi, Sun, Stone,
  Tremblay, Hogan, Eckert, Ettori, Yu, Biffi, \&
  Planelles}]{gaspari_shaken_2018}
Gaspari, M., {McDonald}, M., Hamer, S.~L., {et~al.} 2018, ApJ, 854, 167

\bibitem[{Ghizzardi {et~al.}(2010)Ghizzardi, Rossetti, \&
  Molendi}]{ghizzardi_cold_2010}
Ghizzardi, S., Rossetti, M., \& Molendi, S. 2010, A\&A, 516, A32

\bibitem[{Giacintucci {et~al.}(2014)Giacintucci, Markevitch, Brunetti,
  {ZuHone}, Venturi, Mazzotta, \& Bourdin}]{giacintucci_mapping_2014}
Giacintucci, S., Markevitch, M., Brunetti, G., {et~al.} 2014, ApJ, 795, 73

\bibitem[{Giacintucci {et~al.}(2019)Giacintucci, Markevitch, Cassano, Venturi,
  Clarke, Kale, \& Cuciti}]{giacintucci_expanding_2019}
Giacintucci, S., Markevitch, M., Cassano, R., {et~al.} 2019, ApJ, 880, 70

\bibitem[{Gitti {et~al.}(2006)Gitti, Feretti, \&
  Schindler}]{gitti_multifrequency_2006}
Gitti, M., Feretti, L., \& Schindler, S. 2006, A\&A, 448, 853

\bibitem[{Gupta {et~al.}(2016)Gupta, Yuan, Tran, Martizzi, Taylor, \&
  Kewley}]{gupta_radial_2016}
Gupta, A., Yuan, T., Tran, K.-V.~H., {et~al.} 2016, ApJ, 831, 104

\bibitem[{Hamer {et~al.}(2012)Hamer, Edge, Swinbank, Wilman, Russell, Fabian,
  Sanders, \& Salom}]{hamer_relation_2012}
Hamer, S.~L., Edge, A.~C., Swinbank, A.~M., {et~al.} 2012, MNRAS, 421, 3409

\bibitem[{Hines \& Wills(1993)}]{hines_highly_1993}
Hines, D.~C., \& Wills, B.~J. 1993, ApJ, 415, 82

\bibitem[{{Hitomi Collaboration} {et~al.}(2018){Hitomi Collaboration},
  Aharonian, Akamatsu, Akimoto, Allen, Angelini, Audard, Awaki, Axelsson,
  Bamba, Bautz, Blandford, Brenneman, Brown, Bulbul, Cackett, Canning,
  Chernyakova, Chiao, Coppi, Costantini, de~Plaa, de~Vries, den Herder, Done,
  Dotani, Ebisawa, Eckart, Enoto, Ezoe, Fabian, Ferrigno, Foster, Fujimoto,
  Fukazawa, Furuzawa, Galeazzi, Gallo, Gandhi, Giustini, Goldwurm, Gu,
  Guainazzi, Haba, Hagino, Hamaguchi, Harrus, Hatsukade, Hayashi, Hayashi,
  Hayashi, Hayashida, Hiraga, Hornschemeier, Hoshino, Hughes, Ichinohe, Iizuka,
  Inoue, Inoue, Inoue, Ishida, Ishikawa, Ishisaki, Iwai, Kaastra, Kallman,
  Kamae, Kataoka, Katsuda, Kawai, Kelley, Kilbourne, Kitaguchi, Kitamoto,
  Kitayama, Kohmura, Kokubun, Koyama, Koyama, Kretschmar, Krimm, Kubota,
  Kunieda, Laurent, Lee, Leutenegger, Limousin, Loewenstein, Long, Lumb,
  Madejski, Maeda, Maier, Makishima, Markevitch, Matsumoto, Matsushita,
  McCammon, McNamara, Mehdipour, Miller, Miller, Mineshige, Mitsuda, Mitsuishi,
  Miyazawa, Mizuno, Mori, Mori, Mukai, Murakami, Mushotzky, Nakagawa, Nakajima,
  Nakamori, Nakashima, Nakazawa, Nobukawa, Nobukawa, Noda, Odaka, Ohashi, Ohno,
  Okajima, Ota, Ozaki, Paerels, Paltani, Petre, Pinto, Porter, Pottschmidt,
  Reynolds, Safi-Harb, Saito, Sakai, Sasaki, Sato, Sato, Sato, Sawada,
  Schartel, Serlemtsos, Seta, Shidatsu, Simionescu, Smith, Soong, Stawarz,
  Sugawara, Sugita, Szymkowiak, Tajima, Takahashi, Takahashi, Takeda, Takei,
  Tamagawa, Tamura, Tanaka, Tanaka, Tanaka, Tanaka, Tashiro, Tawara, Terada,
  Terashima, Tombesi, Tomida, Tsuboi, Tsujimoto, Tsunemi, Tsuru, Uchida,
  Uchiyama, Uchiyama, Ueda, Ueda, Uno, Urry, Ursino, Wang, Watanabe, Werner,
  Wilkins, Williams, Yamada, Yamaguchi, Yamaoka, Yamasaki, Yamauchi, Yamauchi,
  Yaqoob, Yatsu, Yonetoku, Zhuravleva, \&
  Zoghbi}]{hitomi_collaboration_atmospheric_2018}
{Hitomi Collaboration}, Aharonian, F., Akamatsu, H., {et~al.} 2018,
  Publications of the Astronomical Society of Japan, 70, 9

\bibitem[{Hlavacek-Larrondo {et~al.}(2013{\natexlab{a}})Hlavacek-Larrondo,
  Fabian, Edge, Ebeling, Allen, Sanders, \&
  Taylor}]{hlavacek-larrondo_rapid_2013}
Hlavacek-Larrondo, J., Fabian, A.~C., Edge, A.~C., {et~al.} 2013{\natexlab{a}},
  MNRAS, 431, 1638

\bibitem[{Hlavacek-Larrondo {et~al.}(2012{\natexlab{a}})Hlavacek-Larrondo,
  Fabian, Edge, Ebeling, Sanders, Hogan, \&
  Taylor}]{hlavacek-larrondo_extreme_2012}
---. 2012{\natexlab{a}}, MNRAS, 421, 1360

\bibitem[{Hlavacek-Larrondo {et~al.}(2012{\natexlab{b}})Hlavacek-Larrondo,
  Fabian, Edge, \& Hogan}]{hlavacek-larrondo_hunt_2012}
Hlavacek-Larrondo, J., Fabian, A.~C., Edge, A.~C., \& Hogan, M.~T.
  2012{\natexlab{b}}, MNRAS, 424, 224

\bibitem[{Hlavacek-Larrondo {et~al.}(2013{\natexlab{b}})Hlavacek-Larrondo,
  Allen, Taylor, Fabian, Canning, Werner, Sanders, Grimes, Ehlert, \& von~der
  Linden}]{hlavacek-larrondo_probing_2013}
Hlavacek-Larrondo, J., Allen, S.~W., Taylor, G.~B., {et~al.}
  2013{\natexlab{b}}, ApJ, 777, 163

\bibitem[{Hlavacek-Larrondo {et~al.}(2015)Hlavacek-Larrondo, {McDonald},
  Benson, Forman, Allen, Bleem, Ashby, Bocquet, Brodwin, Dietrich, Jones, Liu,
  Reichardt, Saliwanchik, Saro, Schrabback, Song, Stalder, Vikhlinin, \&
  Zenteno}]{hlavacek-larrondo_x-ray_2015}
Hlavacek-Larrondo, J., {McDonald}, M., Benson, B.~A., {et~al.} 2015, ApJ, 805,
  35

\bibitem[{Hogan {et~al.}(2015)Hogan, Edge, Hlavacek-Larrondo, Grainge, Hamer,
  Mahony, Russell, Fabian, {McNamara}, \& Wilman}]{hogan_comprehensive_2015}
Hogan, M.~T., Edge, A.~C., Hlavacek-Larrondo, J., {et~al.} 2015, MNRAS, 453,
  1201

\bibitem[{Hogan {et~al.}(2017)Hogan, {McNamara}, Pulido, Nulsen, Vantyghem,
  Russell, Edge, Babyk, Main, \& {McDonald}}]{hogan_onset_2017}
Hogan, M.~T., {McNamara}, B.~R., Pulido, F.~A., {et~al.} 2017, ApJ, 851, 66

\bibitem[{Hopkins {et~al.}(2006)Hopkins, Hernquist, Cox, Di~Matteo, Robertson,
  \& Springel}]{hopkins_unified_2006}
Hopkins, P.~F., Hernquist, L., Cox, T.~J., {et~al.} 2006, ApJ Supplement
  Series, 163, 1

\bibitem[{Iwasawa {et~al.}(2001)Iwasawa, Fabian, \&
  Ettori}]{iwasawa_chandra_2001}
Iwasawa, K., Fabian, A.~C., \& Ettori, S. 2001, MNRAS, 321, L15

\bibitem[{Jauzac {et~al.}(2012)Jauzac, Jullo, Kneib, Ebeling, Leauthaud, Ma,
  Limousin, Massey, \& Richard}]{jauzac_weak_2012}
Jauzac, M., Jullo, E., Kneib, J.-P., {et~al.} 2012, MNRAS, 426, 3369

\bibitem[{Jetzer {et~al.}(2002)Jetzer, Koch, Piffaretti, Puy, \&
  Schindler}]{jetzer_morphology_2002}
Jetzer, P., Koch, P., Piffaretti, R., Puy, D., \& Schindler, S. 2002,
  {arXiv}:astro-ph/0201421, astro-ph/0201421

\bibitem[{Jin {et~al.}(2006)Jin, Zhang, \& Wu}]{jin_hardness_2006}
Jin, Y.~K., Zhang, S.~N., \& Wu, J.~F. 2006, ApJ, 653, 1566

\bibitem[{Johnstone {et~al.}(1987)Johnstone, Fabian, \&
  Nulsen}]{johnstone_optical_1987}
Johnstone, R.~M., Fabian, A.~C., \& Nulsen, P. E.~J. 1987, MNRAS, 224, 75

\bibitem[{Jones \& Forman(1984)}]{jones_structure_1984}
Jones, C., \& Forman, W. 1984, ApJ, 276, 38

\bibitem[{Kalberla {et~al.}(2005)Kalberla, Burton, Hartmann, Arnal, Bajaja,
  Morras, \& Pöppel}]{kalberla_leiden/argentine/bonn_2005}
Kalberla, P. M.~W., Burton, W.~B., Hartmann, D., {et~al.} 2005, A \& A, 440,
  775

\bibitem[{Kaurov {et~al.}(2019)Kaurov, Dai, Venumadhav, Miralda-Escudé, \&
  Frye}]{kaurov_highly_2019}
Kaurov, A.~A., Dai, L., Venumadhav, T., Miralda-Escudé, J., \& Frye, B. 2019,
  ApJ, 880, 58

\bibitem[{Kirkpatrick \& {McNamara}(2015)}]{kirkpatrick_hot_2015}
Kirkpatrick, C.~C., \& {McNamara}, B.~R. 2015, MNRAS, 452, 4361

\bibitem[{Kleinmann {et~al.}(1988)Kleinmann, Hamilton, Keel, Wynn-Williams,
  Eales, Becklin, \& Kuntz}]{kleinmann_properties_1988}
Kleinmann, S.~G., Hamilton, D., Keel, W.~C., {et~al.} 1988, ApJ, 328, 161

\bibitem[{Klesman \& Sarajedini(2012)}]{klesman_multi-wavelength_2012}
Klesman, A.~J., \& Sarajedini, V.~L. 2012, MNRAS, 425, 1215

\bibitem[{Kolokythas {et~al.}(2018)Kolokythas, O'Sullivan, Raychaudhury,
  Giacintucci, Gitti, \& Babul}]{kolokythas_complete_2018}
Kolokythas, K., O'Sullivan, E., Raychaudhury, S., {et~al.} 2018, MNRAS,
  1807.11095

\bibitem[{Li \& Bryan(2014{\natexlab{a}})}]{li_modeling_2014}
Li, Y., \& Bryan, G.~L. 2014{\natexlab{a}}, ApJ, 789, 153

\bibitem[{Li \& Bryan(2014{\natexlab{b}})}]{li_modeling_2014-1}
---. 2014{\natexlab{b}}, ApJ, 789, 54

\bibitem[{Limousin {et~al.}(2012)Limousin, Ebeling, Richard, Swinbank, Smith,
  Jauzac, Rodionov, Ma, Smail, Edge, Jullo, \& Kneib}]{limousin_strong_2012}
Limousin, M., Ebeling, H., Richard, J., {et~al.} 2012, A \& A, 544, A71

\bibitem[{Markevitch \& Vikhlinin(2007)}]{markevitch_shocks_2007}
Markevitch, M., \& Vikhlinin, A. 2007, Physics Reports, 443, 1

\bibitem[{Markevitch {et~al.}(2001)Markevitch, Vikhlinin, \&
  Mazzotta}]{markevitch_non-hydrostatic_2001}
Markevitch, M., Vikhlinin, A., \& Mazzotta, P. 2001, ApJ, 562, L153

\bibitem[{Mazzotta {et~al.}(2003)Mazzotta, Edge, \&
  Markevitch}]{mazzotta_chandra_2003}
Mazzotta, P., Edge, A., \& Markevitch, M. 2003, {ApJ}, 596, 190

\bibitem[{Mazzotta \& Giacintucci(2008)}]{mazzotta_radio_2008}
Mazzotta, P., \& Giacintucci, S. 2008, ApJ Letters, 675, L9

\bibitem[{Mazzotta {et~al.}(2001)Mazzotta, Markevitch, Vikhlinin, \&
  Forman}]{mazzotta_chandra_2001}
Mazzotta, P., Markevitch, M., Vikhlinin, A., \& Forman, W.~R. 2001,
  {arXiv}:astro-ph/0109420, astro-ph/0109420

\bibitem[{{McConnell} {et~al.}(2011){McConnell}, Ma, Gebhardt, Wright, Murphy,
  Lauer, Graham, \& Richstone}]{mcconnell_two_2011}
{McConnell}, N.~J., Ma, C.-P., Gebhardt, K., {et~al.} 2011, Nature, 480, 215

\bibitem[{{McCourt} {et~al.}(2012){McCourt}, Sharma, Quataert, \&
  Parrish}]{mccourt_thermal_2012}
{McCourt}, M., Sharma, P., Quataert, E., \& Parrish, I.~J. 2012, MNRAS, 419,
  3319

\bibitem[{{McDonald} {et~al.}(2018){McDonald}, Gaspari, {McNamara}, \&
  Tremblay}]{mcdonald_revisiting_2018}
{McDonald}, M., Gaspari, M., {McNamara}, B.~R., \& Tremblay, G.~R. 2018, ApJ,
  858, 45

\bibitem[{{McDonald} {et~al.}(2015){McDonald}, {McNamara}, van Weeren,
  Applegate, Bayliss, Bautz, Benson, Carlstrom, Bleem, Chatzikos, Edge, Fabian,
  Garmire, Hlavacek-Larrondo, Jones-Forman, Mantz, Miller, Stalder, Veilleux,
  \& {ZuHone}}]{mcdonald_deep_2015}
{McDonald}, M., {McNamara}, B.~R., van Weeren, R.~J., {et~al.} 2015, ApJ, 811,
  111

\bibitem[{{McDonald} {et~al.}(2019){McDonald}, {McNamara}, Voit, Bayliss,
  Benson, Brodwin, Canning, Florian, Garmire, Gaspari, Gladders,
  Hlavacek-Larrondo, Kara, Reichardt, Russell, Saro, Sharon, Somboonpanyakul,
  Tremblay, \& van Weeren}]{mcdonald_anatomy_2019}
{McDonald}, M., {McNamara}, B.~R., Voit, G.~M., {et~al.} 2019, {arXiv}
  e-prints, 1904, arXiv:1904.08942

\bibitem[{{McNamara} {et~al.}(2016){McNamara}, Russell, Nulsen, Hogan, Fabian,
  Pulido, \& Edge}]{mcnamara_mechanism_2016}
{McNamara}, B.~R., Russell, H.~R., Nulsen, P. E.~J., {et~al.} 2016, ApJ, 830,
  79

\bibitem[{{McNamara} {et~al.}(2000){McNamara}, Wise, Nulsen, David, Sarazin,
  Bautz, Markevitch, Vikhlinin, Forman, Jones, \&
  Harris}]{mcnamara_chandra_2000}
{McNamara}, B.~R., Wise, M., Nulsen, P. E.~J., {et~al.} 2000, ApJ, 534, L135

\bibitem[{{McNamara} {et~al.}(2001){McNamara}, Wise, Nulsen, David, Carilli,
  Sarazin, O'Dea, Houck, Donahue, Baum, Voit, O'Connell, \&
  Koekemoer}]{mcnamara_discovery_2001}
{McNamara}, B.~R., Wise, M.~W., Nulsen, P. E.~J., {et~al.} 2001, ApJ Letters,
  562, L149

\bibitem[{Mushotzky \& Szymkowiak(1988)}]{mushotzky_einstein_1988}
Mushotzky, R.~F., \& Szymkowiak, A.~E. 1988, 229, 53

\bibitem[{Nulsen {et~al.}(2009)Nulsen, Jones, Forman, Churazov, {McNamara},
  David, \& Murray}]{nulsen_radio_2009}
Nulsen, P., Jones, C., Forman, W., {et~al.} 2009, {arXiv}:0909.1809 [astro-ph],
  198

\bibitem[{Nulsen {et~al.}(1987)Nulsen, Johnstone, \& Fabian}]{nulsen_star_1987}
Nulsen, P. E.~J., Johnstone, R.~M., \& Fabian, A.~C. 1987, Proceedings of the
  Astronomical Society of Australia, 7, 132

\bibitem[{Nulsen \& {McNamara}(2013)}]{nulsen_agn_2013}
Nulsen, P. E.~J., \& {McNamara}, B.~R. 2013, Astron. Nachr., 334, 386

\bibitem[{O'Dea {et~al.}(2008)O'Dea, Baum, Privon, Noel-Storr, Quillen, Zufelt,
  Park, Edge, Russell, Fabian, Donahue, Sarazin, {McNamara}, Bregman, \&
  Egami}]{odea_infrared_2008}
O'Dea, C.~P., Baum, S.~A., Privon, G., {et~al.} 2008, ApJ, 681, 1035

\bibitem[{O'Sullivan {et~al.}(2011)O'Sullivan, Giacintucci, David, Gitti,
  Vrtilek, Raychaudhury, \& Ponman}]{osullivan_heating_2011}
O'Sullivan, E., Giacintucci, S., David, L.~P., {et~al.} 2011, ApJ, 735, 11

\bibitem[{Owers {et~al.}(2009)Owers, Nulsen, Couch, \&
  Markevitch}]{owers_high_2009}
Owers, M.~S., Nulsen, P. E.~J., Couch, W.~J., \& Markevitch, M. 2009, {ApJ},
  704, 1349

\bibitem[{Panagoulia {et~al.}(2014)Panagoulia, Fabian, \&
  Sanders}]{panagoulia_volume-limited_2014}
Panagoulia, E.~K., Fabian, A.~C., \& Sanders, J.~S. 2014, MNRAS, 438, 2341

\bibitem[{Peterson \& Fabian(2006)}]{peterson_x-ray_2006}
Peterson, J.~R., \& Fabian, A.~C. 2006, Physics Reports, 427, 1

\bibitem[{Pfrommer \& Enßlin(2004)}]{pfrommer_constraining_2004}
Pfrommer, C., \& Enßlin, T.~A. 2004, A \& A, 413, 17

\bibitem[{Piconcelli {et~al.}(2007)Piconcelli, Fiore, Nicastro, Mathur, Brusa,
  Comastri, \& Puccetti}]{piconcelli_xmm-newton_2007}
Piconcelli, E., Fiore, F., Nicastro, F., {et~al.} 2007, A \& A, 473, 85

\bibitem[{{Planck Collaboration} {et~al.}(2015){Planck Collaboration}, Ade,
  Aghanim, Armitage-Caplan, Arnaud, Ashdown, Atrio-Barandela, Aumont, Aussel,
  Baccigalupi, Banday, Barreiro, Barrena, Bartelmann, Bartlett, Battaner,
  Benabed, Benoît, Benoit-Lévy, Bernard, Bersanelli, Bielewicz, Bikmaev,
  Bobin, Bock, Böhringer, Bonaldi, Bond, Borrill, Bouchet, Bridges, Bucher,
  Burenin, Burigana, Butler, Cardoso, Carvalho, Catalano, Challinor, Chamballu,
  Chary, Chen, Chiang, Chiang, Chon, Christensen, Churazov, Church, Clements,
  Colombi, Colombo, Comis, Couchot, Coulais, Crill, Curto, Cuttaia, Da~Silva,
  Dahle, Danese, Davies, Davis, de~Bernardis, de~Rosa, de~Zotti, Delabrouille,
  Delouis, Démoclès, Désert, Dickinson, Diego, Dolag, Dole, Donzelli, Doré,
  Douspis, Dupac, Efstathiou, Enßlin, Eriksen, Feroz, Ferragamo, Finelli,
  Flores-Cacho, Forni, Frailis, Franceschi, Fromenteau, Galeotta, Ganga,
  Génova-Santos, Giard, Giardino, Gilfanov, Giraud-Héraud, González-Nuevo,
  Górski, Grainge, Gratton, Gregorio, Groeneboom, Gruppuso, Hansen, Hanson,
  Harrison, Hempel, Henrot-Versillé, Hernández-Monteagudo, Herranz,
  Hildebrandt, Hivon, Hobson, Holmes, Hornstrup, Hovest, Huffenberger, Hurier,
  Hurley-Walker, Jaffe, Jaffe, Jones, Juvela, Keihänen, Keskitalo, Khamitov,
  Kisner, Kneissl, Knoche, Knox, Kunz, Kurki-Suonio, Lagache, Lähteenmäki,
  Lamarre, Lasenby, Laureijs, Lawrence, Leahy, Leonardi, León-Tavares,
  Lesgourgues, Li, Liddle, Liguori, Lilje, Linden-Vørnle, López-Caniego,
  Lubin, Macías-Pérez, MacTavish, Maffei, Maino, Mandolesi, Maris, Marshall,
  Martin, Martínez-González, Masi, Massardi, Matarrese, Matthai, Mazzotta,
  Mei, Meinhold, Melchiorri, Melin, Mendes, Mennella, Migliaccio, Mikkelsen,
  Mitra, Miville-Deschênes, Moneti, Montier, Morgante, Mortlock, Munshi,
  Murphy, Naselsky, Nastasi, Nati, Natoli, Nesvadba, Netterfield,
  Nørgaard-Nielsen, Noviello, Novikov, Novikov, O'Dwyer, Olamaie, Osborne,
  Oxborrow, Paci, Pagano, Pajot, Paoletti, Pasian, Patanchon, Pearson,
  Perdereau, Perotto, Perrott, Perrotta, Piacentini, Piat, Pierpaoli,
  Pietrobon, Plaszczynski, Pointecouteau, Polenta, Ponthieu, Popa, Poutanen,
  Pratt, Prézeau, Prunet, Puget, Rachen, Reach, Rebolo, Reinecke, Remazeilles,
  Renault, Ricciardi, Riller, Ristorcelli, Rocha, Rosset, Roudier,
  Rowan-Robinson, Rubiño-Martín, Rumsey, Rusholme, Sandri, Santos, Saunders,
  Savini, Schammel, Scott, Seiffert, Shellard, Shimwell, Spencer, Starck,
  Stolyarov, Stompor, Streblyanska, Sudiwala, Sunyaev, Sureau, Sutton,
  Suur-Uski, Sygnet, Tauber, Tavagnacco, Terenzi, Toffolatti, Tomasi, Tramonte,
  Tristram, Tucci, Tuovinen, Türler, Umana, Valenziano, Valiviita, Van~Tent,
  Vibert, Vielva, Villa, Vittorio, Wade, Wandelt, White, White, Yvon, Zacchei,
  \& Zonca}]{planck_collaboration_planck_2015}
{Planck Collaboration}, Ade, P. A.~R., Aghanim, N., {et~al.} 2015, A \& A, 581,
  A14

\bibitem[{Planck~Collaboration {et~al.}(2011)Planck~Collaboration, Ade,
  Aghanim, Arnaud, Ashdown, Aumont, Baccigalupi, Balbi, Banday, Barreiro,
  Bartelmann, Bartlett, Battaner, Benabed, Benoît, Bernard, Bersanelli,
  Bhatia, Bock, Bonaldi, Bond, Borrill, Bouchet, Bourdin, Brown, Bucher,
  Burigana, Cabella, Cardoso, Catalano, Cayón, Challinor, Chamballu, Chiang,
  Chiang, Chon, Christensen, Churazov, Clements, Colafrancesco, Colombi,
  Couchot, Coulais, Crill, Cuttaia, Silva, Dahle, Danese, Bernardis, Gasperis,
  Rosa, Zotti, Delabrouille, Delouis, Désert, Diego, Dolag, Donzelli, Doré,
  Dörl, Douspis, Dupac, Efstathiou, Enßlin, Finelli, Flores-Cacho, Forni,
  Frailis, Franceschi, Fromenteau, Galeotta, Ganga, Génova-Santos, Giard,
  Giardino, Giraud-Héraud, González-Nuevo, Górski, Gratton, Gregorio,
  Gruppuso, Harrison, Henrot-Versillé, Hernández-Monteagudo, Herranz,
  Hildebrandt, Hivon, Hobson, Holmes, Hovest, Hoyland, Huffenberger, Jaffe,
  Jones, Juvela, Keihänen, Keskitalo, Kisner, Kneissl, Knox, Kurki-Suonio,
  Lagache, Lamarre, Lanoux, Lasenby, Laureijs, Lawrence, Leach, Leonardi,
  Liddle, Lilje, Linden-Vørnle, López-Caniego, Lubin, Macías-Pérez,
  {MacTavish}, Maffei, Maino, Mandolesi, Mann, Maris, Marleau,
  Martínez-González, Masi, Matarrese, Matthai, Mazzotta, Melchiorri, Melin,
  Mendes, Mennella, Mitra, Miville-Deschênes, Moneti, Montier, Morgante,
  Mortlock, Munshi, Murphy, Naselsky, Natoli, Netterfield, Nørgaard-Nielsen,
  Noviello, Novikov, Novikov, Osborne, Pajot, Pasian, Patanchon, Perdereau,
  Perotto, Perrotta, Piacentini, Piat, Pierpaoli, Piffaretti, Plaszczynski,
  Pointecouteau, Polenta, Ponthieu, Poutanen, Pratt, Prézeau, Prunet, Puget,
  Rachen, Rebolo, Reinecke, Renault, Ricciardi, Riller, Ristorcelli, Rocha,
  Rosset, Rubiño-Martín, Rusholme, Sandri, Santos, Savini, Schaefer, Scott,
  Seiffert, Shellard, Smoot, Starck, Stivoli, Stolyarov, Sudiwala, Sunyaev,
  Sygnet, Tauber, Terenzi, Toffolatti, Tomasi, Torre, Tristram, Tuovinen,
  Valenziano, Vibert, Vielva, Villa, Vittorio, Wade, Wandelt, White, White,
  Yvon, Zacchei, \& Zonca}]{ade_planck_2011}
Planck~Collaboration, C., Ade, P. a.~R., Aghanim, N., {et~al.} 2011, A\&A, 536,
  A11

\bibitem[{Prasad {et~al.}(2017)Prasad, Sharma, \& Babul}]{prasad_agn_2017}
Prasad, D., Sharma, P., \& Babul, A. 2017, MNRAS, 471, 1531

\bibitem[{Rafferty {et~al.}(2008)Rafferty, {McNamara}, \&
  Nulsen}]{rafferty_regulation_2008}
Rafferty, D.~A., {McNamara}, B.~R., \& Nulsen, P. E.~J. 2008, ApJ, 687, 899

\bibitem[{Rafferty {et~al.}(2006)Rafferty, {McNamara}, Nulsen, \&
  Wise}]{rafferty_feedback-regulated_2006}
Rafferty, D.~A., {McNamara}, B.~R., Nulsen, P. E.~J., \& Wise, M.~W. 2006, ApJ,
  652, 216

\bibitem[{Richard-Laferri\`{e}re {et~al.}(submitted)Richard-Laferri\`{e}re,
  Hlavacek-Larrondo, Latulippe, Nemmen, Prasow-\'{E}mond, Taylor, Edge, Rhea,
  Fabian, Gendron-Marsolais, Hogan, Sanders, \& Demontigny}]{richard}
Richard-Laferri\`{e}re, A., Hlavacek-Larrondo, J., Latulippe, M., {et~al.}
  submitted, MNRAS

\bibitem[{Russell {et~al.}(2013)Russell, {McNamara}, Edge, Hogan, Main, \&
  Vantyghem}]{russell_radiative_2013}
Russell, H.~R., {McNamara}, B.~R., Edge, A.~C., {et~al.} 2013, MNRAS, 432, 530

\bibitem[{Russell {et~al.}(2012)Russell, {McNamara}, Sanders, Fabian, Nulsen,
  Canning, Baum, Donahue, Edge, King, \& O'Dea}]{russell_shock_2012}
Russell, H.~R., {McNamara}, B.~R., Sanders, J.~S., {et~al.} 2012, MNRAS, 423,
  236

\bibitem[{Sanders \& Fabian(2002)}]{sanders_spatially_2002}
Sanders, J.~S., \& Fabian, A.~C. 2002, MNRAS, 331, 273

\bibitem[{Sanders \& Fabian(2007)}]{sanders_deeper_2007}
---. 2007, MNRAS, 381

\bibitem[{Sanders {et~al.}(2018)Sanders, Fabian, Russell, \&
  Walker}]{sanders_hydrostatic_2018}
Sanders, J.~S., Fabian, A.~C., Russell, H.~R., \& Walker, S.~A. 2018, MNRAS,
  474, 1065

\bibitem[{Sanders {et~al.}(2016)Sanders, Fabian, Taylor, Russell, Blundell,
  Canning, Hlavacek-Larrondo, Walker, \& Grimes}]{sanders_very_2016}
Sanders, J.~S., Fabian, A.~C., Taylor, G.~B., {et~al.} 2016, Mon. Not. R.
  Astron. Soc., 457, 82

\bibitem[{Santos {et~al.}(2016)Santos, Balestra, Tozzi, Altieri, Valtchanov,
  Mercurio, Nonino, Yu, Rosati, Grillo, Medezinski, \&
  Biviano}]{santos_starbursting_2016}
Santos, J.~S., Balestra, I., Tozzi, P., {et~al.} 2016, Mon. Not. R. Astron.
  Soc: Lett., 456, L99

\bibitem[{Schindler {et~al.}(2001)Schindler, Castillo-Morales, De~Filippis,
  Schwope, \& Wambsganss}]{schindler_discovery_2001}
Schindler, S., Castillo-Morales, A., De~Filippis, E., Schwope, A., \&
  Wambsganss, J. 2001, A\&A, 376, L27

\bibitem[{Sharma {et~al.}(2012)Sharma, {McCourt}, Quataert, \&
  Parrish}]{sharma_thermal_2012}
Sharma, P., {McCourt}, M., Quataert, E., \& Parrish, I.~J. 2012, MNRAS, 420,
  3174

\bibitem[{Simionescu {et~al.}(2010)Simionescu, Werner, Forman, Miller, Takei,
  Böhringer, Churazov, \& Nulsen}]{simionescu_metal_2010}
Simionescu, A., Werner, N., Forman, W.~R., {et~al.} 2010, MNRAS, 405, 91

\bibitem[{Smith {et~al.}(2009)Smith, Ebeling, Limousin, Kneib, Swinbank, Ma,
  Jauzac, Richard, Jullo, Sand, Edge, \& Smail}]{smith_hubble_2009}
Smith, G.~P., Ebeling, H., Limousin, M., {et~al.} 2009, ApJ Letters, 707, L163

\bibitem[{Su {et~al.}(2017)Su, Nulsen, Kraft, Forman, Jones, Irwin, Randall, \&
  Churazov}]{su_buoyant_2017}
Su, Y., Nulsen, P. E.~J., Kraft, R.~P., {et~al.} 2017, ApJ, 847, 94

\bibitem[{van Weeren {et~al.}(2019)van Weeren, de~Gasperin, Akamatsu, Brüggen,
  Feretti, Kang, Stroe, \& Zandanel}]{van_weeren_diffuse_2019}
van Weeren, R.~J., de~Gasperin, F., Akamatsu, H., {et~al.} 2019, Space Science
  Reviews, 215, 16

\bibitem[{Vignali {et~al.}(2011)Vignali, Piconcelli, Lanzuisi, Feltre,
  Feruglio, Maiolino, Fiore, Fritz, La~Parola, Mignoli, \&
  Pozzi}]{vignali_nature_2011}
Vignali, C., Piconcelli, E., Lanzuisi, G., {et~al.} 2011, MNRAS, 416, 2068

\bibitem[{Vikhlinin {et~al.}(2005)Vikhlinin, Markevitch, Murray, Jones, Forman,
  \& Van~Speybroeck}]{vikhlinin_chandra_2005}
Vikhlinin, A., Markevitch, M., Murray, S.~S., {et~al.} 2005, ApJ, 628, 655

\bibitem[{Voigt \& Fabian(2004)}]{voigt_thermal_2004}
Voigt, L.~M., \& Fabian, A.~C. 2004, MNRAS, 347, 1130

\bibitem[{Voit {et~al.}(2017)Voit, Meece, Li, O'Shea, Bryan, \&
  Donahue}]{voit_global_2017}
Voit, G.~M., Meece, G., Li, Y., {et~al.} 2017, ApJ, 845, 80

\bibitem[{von~der Linden {et~al.}(2014)von~der Linden, Mantz, Allen, Applegate,
  Kelly, Morris, Wright, Allen, Burchat, Burke, Donovan, \&
  Ebeling}]{von_der_linden_robust_2014}
von~der Linden, A., Mantz, A., Allen, S.~W., {et~al.} 2014, MNRAS, 443, 1973

\bibitem[{Walker {et~al.}(2014)Walker, Fabian, \&
  Kosec}]{walker_exploring_2014}
Walker, S.~A., Fabian, A.~C., \& Kosec, P. 2014, MNRAS, 445, 3444

\bibitem[{Walker {et~al.}(2015)Walker, Kosec, Fabian, \&
  Sanders}]{walker_x-ray_2015}
Walker, S.~A., Kosec, P., Fabian, A.~C., \& Sanders, J.~S. 2015, Mon. Not. R.
  Astron. Soc., 453, 2481

\bibitem[{Wang {et~al.}(2004)Wang, Malhotra, Rhoads, \&
  Norman}]{wang_identifying_2004}
Wang, J.~X., Malhotra, S., Rhoads, J.~E., \& Norman, C.~A. 2004, ApJ, 612, L109

\bibitem[{Zandanel {et~al.}(2014)Zandanel, Pfrommer, \&
  Prada}]{zandanel_physics_2014}
Zandanel, F., Pfrommer, C., \& Prada, F. 2014, MNRAS, 438, 124

\bibitem[{Zhuravleva {et~al.}(2018)Zhuravleva, Allen, Mantz, \&
  Werner}]{zhuravleva_gas_2018}
Zhuravleva, I., Allen, S.~W., Mantz, A., \& Werner, N. 2018, ApJ, 865, 53

\bibitem[{Zhuravleva {et~al.}(2014)Zhuravleva, Churazov, Schekochihin, Allen,
  Arévalo, Fabian, Forman, Sanders, Simionescu, Sunyaev, Vikhlinin, \&
  Werner}]{zhuravleva_turbulent_2014}
Zhuravleva, I., Churazov, E., Schekochihin, A.~A., {et~al.} 2014, Nature, 515,
  85

\bibitem[{Zhuravleva {et~al.}(2016)Zhuravleva, Churazov, Arévalo,
  Schekochihin, Forman, Allen, Simionescu, Sunyaev, Vikhlinin, \&
  Werner}]{zhuravleva_nature_2016}
Zhuravleva, I., Churazov, E., Arévalo, P., {et~al.} 2016, MNRAS, 458, 2902

\bibitem[{Zitrin {et~al.}(2009)Zitrin, Broadhurst, Rephaeli, \&
  Sadeh}]{zitrin_largest_2009}
Zitrin, A., Broadhurst, T., Rephaeli, Y., \& Sadeh, S. 2009, ApJ Letters, 707,
  L102

\bibitem[{{ZuHone} {et~al.}(2013){ZuHone}, Markevitch, Brunetti, \&
  Giacintucci}]{zuhone_turbulence_2013}
{ZuHone}, J.~A., Markevitch, M., Brunetti, G., \& Giacintucci, S. 2013, ApJ,
  762, 78

\end{thebibliography}

\end{document}